\pdfoutput=1
\documentclass[journal]{IEEEtran}
\usepackage{amsmath,amssymb,amsfonts}
\usepackage{float} 
\usepackage{graphicx}
\graphicspath{{figures/}}
\usepackage{algorithm}
\usepackage{algorithmic} 
\usepackage{booktabs}
\usepackage{siunitx}
\usepackage{xcolor}
\usepackage[caption=false,font=footnotesize]{subfig}
\begin{document}

\title{Inter-Cell Interference Suppression for Channel Estimation and Data Detection in MIMO-OFDM Systems via Score-Based Variational Inference}
\author{Mingyang Yue
    and Xiaojun Yuan,~\IEEEmembership{Fellow,~IEEE}% <-this % stops a space
\thanks{This work has been submitted to the IEEE for possible publication. Copyright may be transferred without notice, after which this version may no longer be accessible.}
\thanks{
    This work was supported in part by the National Natural Science Foundation of China under Grant 62571087, and in part by the National Key Laboratory of Wireless Communications Foundation under Grant IFN202507. A preliminary version of this work was published in the 11th International Conference on Computer and Communication Systems (ICCCS)~\cite{yue_icccs2026}.
    
Mingyang Yue and Xiaojun Yuan are with the National  Key Laboratory of Wireless Communications, University of Electronic Science and Technology of China, Chengdu 611731, China (e-mail:myyue@std.uestc.edu.cn; xjyuan@uestc.edu.cn).}}

\maketitle

\begin{abstract}
In multiple-input multiple-output orthogonal frequency-division multiplexing (MIMO-OFDM) systems, conventional receivers based on linear channel estimation and subband-averaged interference covariance reconstruction suffer serious performance degradation under strong interference. To address this issue, we propose a probabilistic framework for joint channel estimation and interference suppression using score-based prior learning. Specifically, we formulate a variational inference problem that exploits the bilinear structure of the interference by explicitly separating the interference channels from the interference symbols. To solve the resulting highly nonconvex bilinear inference problem, we develop a score-based variational inference  algorithm that combines score-based turbo message passing  for target-channel estimation with a reverse mean propagation framework for interference-channel inference, while the interference symbols are updated via linear minimum mean square error (LMMSE) estimation. We further design an interference rejection combining (IRC) detector with subcarrier-level covariance reconstruction based on the estimated interference channels, thereby avoiding the performance loss caused by conventional subband averaging. Numerical results demonstrate that the proposed algorithm is able to effectively suppress the interference in channel estimation, as well as to accurately reconstruct the interference subspace for interference rejection in data detection, and thus significantly outperforms existing baselines.
\end{abstract}
    

\section{Introduction}\label{sec:intro}
Multiple-input multiple-output orthogonal frequency-division multiplexing (MIMO-OFDM) is a cornerstone technology in modern wireless communications, owing to its high spectral efficiency, spatial multiplexing capability, and robustness over frequency-selective fading channels~\cite{tse2005fundamentals}. 
These advantages, however, come at the cost of increased receiver complexity. 
In MIMO-OFDM systems, channel estimation (CE) is challenging because the wireless channel is typically high-dimensional and exhibits complicated spatial-frequency variations, while accurate channel state information (CSI) acquisition is further constrained by limited pilot resources and noise contamination. 
Data detection is also challenging, since the receiver must recover multiple transmitted streams from superimposed observations, often under severe inter-stream interference, channel uncertainty, and practical complexity constraints. 

Extensive research has been devoted to CE for MIMO-OFDM systems. 
Linear estimators such as the linear minimum mean square error (LMMSE) estimator are widely used in practice because of their low complexity and analytical tractability when reliable second-order channel statistics are available. 
Methods based on explicit or closed-form mathematical priors exploit structural properties of wireless channels, such as angle--delay-domain sparsity, second-order statistics, or Markov structure, to perform channel estimation~\cite{8240714,11310707}. Structured hybrid message-passing methods further combine explicit priors with module-wise inference~\cite{liu2023structured,song2023hybridmp}. Such methods are interpretable and computationally efficient when their modeling assumptions are accurate, but their performance can be limited by restrictive or mismatched prior models.
In parallel, end-to-end data-driven CE methods based on DNN~\cite{8400482}, CNN~\cite{11045666,10534815}, or attention-based~\cite{ARA} architectures learn a direct mapping from received observations to channel estimates from data. Deep-unfolding methods~\cite{9685707,8353153,10075639}, in contrast, embed trainable components or parameters into model-based iterations and therefore constitute a hybrid model- and data-driven approach.
More recently, score-based generative models~\cite{song2021scorebased} learn the score, i.e., the gradient of the log-density, of noise-perturbed data distributions. Rather than directly learning an observation-to-estimate mapping, score-based CE methods~\cite{score_CE1,score_CE2,score_CE3} use the learned score as a generative channel prior in probabilistic inference. This learned prior can capture complex channel distributions that are difficult to specify analytically, while requiring representative training data and neural-network inference and potentially being sensitive to distribution mismatch.
Overall, existing CE
methods have substantially improved the accuracy of CE,
thereby facilitating subsequent data detection.

Data detection in MIMO systems has also been extensively studied. 
Maximum-likelihood detection and sphere decoding~\cite{sd} can approach optimal performance, but their complexity grows rapidly with the system dimension and modulation order. 
Linear detectors, such as the minimum mean-squared error (MMSE) detector, remain attractive in practice because of their low computational cost, although their performance may degrade significantly in challenging propagation environments. 
Approximate message-passing methods~\cite{amp_md} provide scalable iterative alternatives by leveraging prior distributions and probabilistic factorization, but their convergence may become fragile under strong channel correlation or model mismatch. 
More recently, deep-learning-based detectors~\cite{learndetect,oamp,smd,pure} have shown improved flexibility by learning nonlinear decision rules from data. 
Nevertheless, as with many data-driven approaches, their robustness may deteriorate when the operating environment deviates from the training distribution. 
Overall, existing data detection methods offer different tradeoffs among performance, complexity, and robustness.

Despite these advances in CE and data detection, practical MIMO-OFDM receivers often operate in the presence of strong non-cooperative inter-cell co-channel interference, especially in dense cellular deployments where the reuse of time-frequency resources and the increasing number of active users make cell-edge reception interference-limited.
Since adjacent-cell transmissions share the same band and propagate through multipath channels, the received interference can occupy a rich spatial subspace. It is therefore difficult to suppress it using purely frequency or spatial sparsity.
This setting introduces a fundamental challenge to both CE and data detection.  On one hand, the
interferer is generally non-cooperative, and hence neither its
transmitted symbols nor its instantaneous channels are known
to the receiver. That is, the receiver has no side information,
such as the pilot sequences of the interferers, to facilitate
interference-channel estimation. On the other hand, the interference
is typically time- and frequency-varying, and may
exhibit complicated spatial and spectral structures. As a result,
the receiver must estimate the target channel and detect the
transmitted data while facing interference whose statistical
characteristics are difficult to characterize explicitly and whose
realizations vary dynamically across antennas and subcarriers.

It is useful to distinguish the blind setting considered here from  network-assisted interference cancellation. Network-assisted interference cancellation and suppression (NAICS) was studied in 3GPP Release~12 to exploit side information about dominant interferers, such as pilot sequences, modulation formats, transmit powers, and spatial-layer or precoding information~\cite{3gpp_naics}. However, such fully assisted receiver architectures depend on accurate and timely cross-node signaling and have not been deployed widely in practical cellular systems. By contrast, minimum mean-square error interference rejection combining (MMSE-IRC) is a more commonly adopted covariance-based receiver principle for suppressing colored interference and does not require complete instantaneous side information about the interferers~\cite{winters1984optimum,3gpp_mmseirc}. This paper focuses on blind or restricted deployments, where the required interferer side information is unavailable, unreliable, or outdated; the receiver must therefore infer the interference structure from its own received observations.

Most existing CE and data detection methods, however, are
developed under additive isotropic noise, and do not explicitly
exploit the latent structure of such interference.
To date, very limited studies in the literature have considered interference-aware CE and data detection. 
For CE, reference~\cite{DSAE} estimated the spatial correlation of the interference by averaging observations over the delay domain, and then combined LMMSE estimation with autoencoder-based refinement in the sparse angle-delay domain. 
Although effective to a certain extent, such a coarse representation cannot fully capture the fine-grained spatio-frequency variation of dynamic interference. 
In this blind setting, practical MMSE-IRC implementations estimate the interference-plus-noise covariance through subband or resource-block averaging~\cite{3gpp_mmseirc,cho2010mimo}.
Its effectiveness, however, strongly depends on the quality of the interference-plus-noise covariance estimate. 
These sample-average strategies implicitly assume that the interference statistics remain locally stationary, and therefore tend to smooth out the frequency-selective characteristics of the true interference. 
Consequently, existing interference-aware receivers still rely largely on coarse statistical approximations and do not fully exploit the underlying structure of the interference across CE and data detection.

Motivated by the above observations, this paper develops
a probabilistic framework for joint channel estimation and
interference suppression (JCEIS) in MIMO-OFDM systems
under strong interference. The key idea is to decompose
the interference into interference channels and interference
symbols, and to characterize the analytically intractable priors
of the target channel and the interference channels with learned score
priors, so that the latent variables can be inferred
jointly with the target channel. Numerical results demonstrate
that the proposed framework achieves significant gains in both
CE and data detection  in
strongly interference-limited regimes.

The main contributions of this paper are summarized as follows:
\begin{itemize}
    \item We propose a variational inference framework for JCEIS in MIMO-OFDM systems under strong interference. The framework uses score-based priors to characterize analytically intractable target-channel and interference distributions. Instead of modeling aggregate interference directly, it factorizes the interference into interference channels and symbols and uses a score-based prior for the interference channels.

    \item Based on this formulation, we develop the score-based variational-inference algorithm JCEIS-SVI. It integrates a score-based turbo message passing (STMP) module for target-channel estimation with a reverse mean propagation (RMP) module for interference-channel inference, while LMMSE updates the interference symbols.

    \item Based on the estimated interference channels and symbols, we design a projection-based covariance reconstruction scheme for MMSE-IRC detection. The estimated channels identify the interference subspace, from which the covariance is reconstructed at the subcarrier level to improve robustness against frequency-selective interference.
\end{itemize}

Compared with the preliminary conference version~\cite{yue_icccs2026}, this journal version substantially extends the receiver design from channel estimation to joint channel estimation and interference suppression, introduces projection-based covariance reconstruction for MMSE-IRC detection, and provides expanded analysis and simulations under strong non-cooperative interference.

\textit{Notations:} Throughout this paper, scalars, vectors, and matrices are denoted by non-boldface letters, boldface lower-case letters, and boldface upper-case letters, respectively.
The superscripts ${(\cdot)}^{\mathsf T}$ and ${(\cdot)}^{\mathsf H}$ denote the transpose and conjugate-transpose operations, respectively. 
The trace and Frobenius norm of a matrix are denoted by $\operatorname{Tr}(\cdot)$ and $\|\cdot\|_{\text{F}}$, respectively. 
The expectation operator is denoted by $\mathbb{E}[\cdot]$. 
A diagonal matrix with diagonal entries $x_1,\dots,x_n$ is denoted by $\operatorname{diag}\{x_1,\dots,x_n\}$. 
The notation $\mathbf{x}\sim\mathcal{CN}(\boldsymbol{\mu},\mathbf{\Sigma})$ indicates that $\mathbf{x}$ follows a circularly symmetric complex Gaussian (CSCG) distribution. 
The identity matrix with an appropriate dimension is denoted by $\mathbf{I}$, and $\delta(\cdot)$ denotes the Dirac delta function. 
In addition, $\mathcal{VM}(\mu,\kappa)$ and $\mathcal{U}(a,b)$ denote the von Mises and uniform distributions, respectively.

\section{Preliminaries}\label{sec:preliminaries}

\subsection{Channel Model}
\begin{figure}[htbp]
  \centering
  \includegraphics[width=0.45\textwidth]{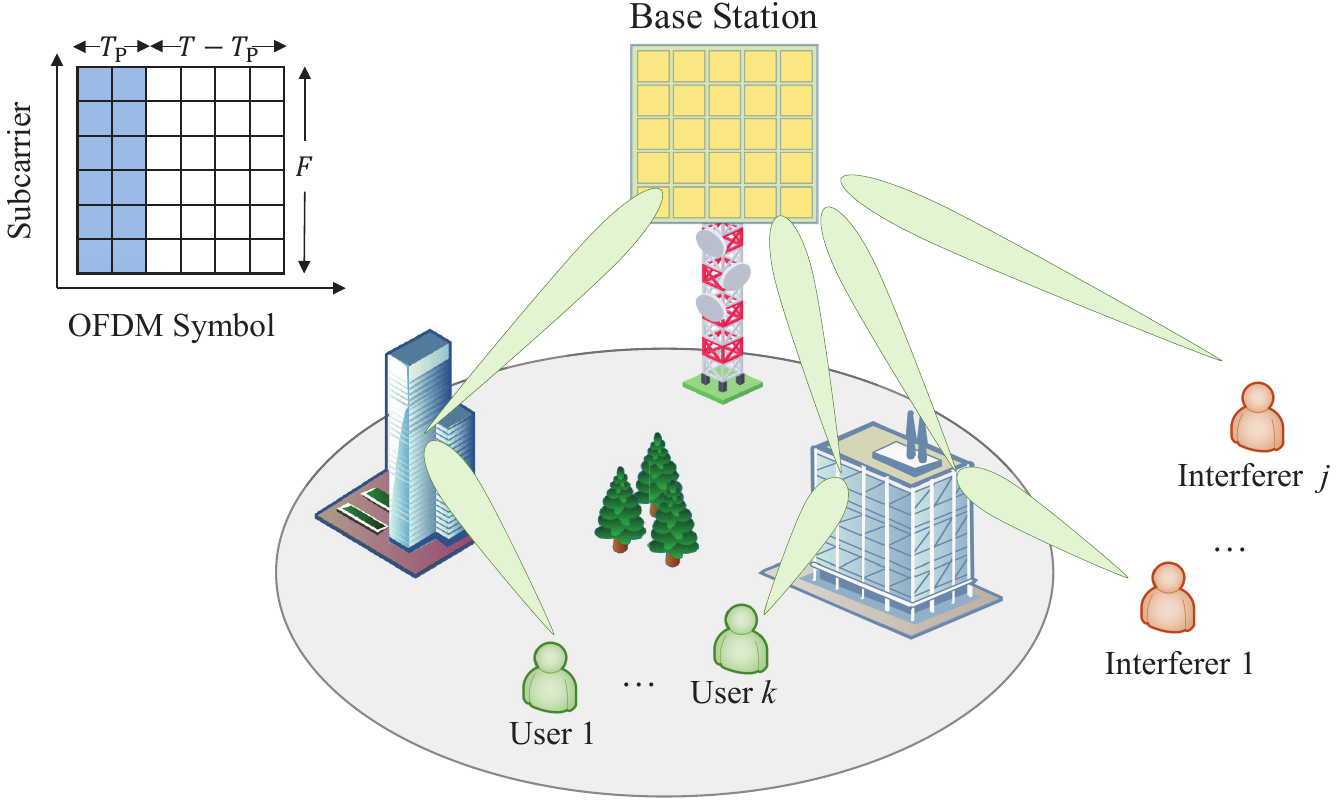}
  \caption{Uplink multi-user MIMO-OFDM system with interference.}\label{fig:system}
\end{figure}
As illustrated in Fig.~\ref{fig:system}, we consider an uplink multi-user MIMO-OFDM communication system, where a base station (BS) equipped with $N$ receive antennas serves $K$ single-antenna users. The system operates in the presence of $J$  interference sources originating from non-cooperative adjacent cells.
Each transmission frame comprises $T$ OFDM symbols and occupies $F$ subcarriers with spacing of $\Delta f$. We assume  a block-fading channel model, i.e.,  the channel remains quasi-static within a frame but varies independently over frames. 

After removing the cyclic prefix and applying the discrete Fourier transform, the received signal vector $\mathbf{y}_{f,t} \in \mathbb{C}^{N \times 1}$ at the $t$-th OFDM symbol and the $f$-th subcarrier is modeled as
\begin{equation}
    \mathbf{y}_{f,t} = \sum_{k=1}^{K} \mathbf{h}_{k,f} x_{k,f,t} + \sum_{j=1}^{J} \mathbf{l}_{j, f, t} + \mathbf{n}_{f,t},
    \label{eq:scalar_model}
\end{equation}
where $\mathbf{h}_{k,f} \in \mathbb{C}^{N \times 1}$ denotes the channel vector for the $k$-th user at the  $f$-th subcarrier, $x_{k,f,t}$  the transmitted symbol of the $k$-th user at the $t$-th OFDM symbol and the $f$-th subcarrier, $\mathbf{l}_{j,f,t} \in \mathbb{C}^{N \times 1}$  the interference signal from the $j$-th interferer at the $t$-th OFDM symbol and the $f$-th subcarrier, and $\mathbf{n}_{f,t} \sim \mathcal{CN}(\mathbf{0},\sigma^2 \mathbf{I})$ the AWGN at the $t$-th OFDM symbol and the $f$-th subcarrier.\@ 
The channel vector $\mathbf{h}_{k,f}$ is characterized by the geometric multipath model~\cite{tse2005fundamentals}, given by
\begin{align}
    \mathbf{h}_{k,f} = \sum_{l=1}^{L_k^{\text{tar}}} \epsilon_{k,l}^{\rm tar}\mathbf{a}_{\mathsf{S}}(\theta_{k,l}^{\rm tar}, \beta_{k,l}^{\rm tar}) e^{-\jmath 2 \pi f \Delta f \tau_{k,l}^{\rm tar} },
\end{align}
where $L_k^{\text{tar}}$ denotes the number of propagation paths of user $k$, and $\epsilon_{k,l}^{\rm tar}$ the complex gain for the $l$-th path with average power $\rho_{k,l}^{\rm tar}$. The parameters $\tau_{k,l}^{\rm tar}$, $\theta_{k,l}^{\rm tar}$, and $\beta_{k,l}^{\rm tar}$ are its delay, azimuth, and elevation AoAs, respectively. The vector $\mathbf{a}_{\mathsf{S}}(\theta, \varphi) \in \mathbb{C}^{N \times 1}$ denotes the spatial steering vector of the array at the BS.\@
By aggregating the channel vectors across all subcarriers, the space-frequency channel matrix for the $k$-th user, $\mathbf{H}_{k}={[\mathbf{h}_{k,1}, \ldots, \mathbf{h}_{k, F}]}^{\mathrm{T}} \in \mathbb{C}^{F \times N}$, can be compactly expressed as
\begin{align}
    \mathbf{H}_{k} = \sum_{l=1}^{L_k^{\text{tar}}} \epsilon_{k,l}^{\rm tar} \mathbf{a}_{\mathsf{F}}(\tau_{k,l}^{\rm tar}) {\mathbf{a}_{\mathsf{S}}(\theta_{k,l}^{\rm tar}, \beta_{k,l}^{\rm tar})}^{\mathrm{T}},
\end{align}
where $\mathbf{a}_{\mathsf{F}}(\tau) = {[e^{-\jmath 2 \pi \Delta f \tau}, \ldots, e^{-\jmath 2 \pi F \Delta f \tau}]}^{\mathrm{T}} \in \mathbb{C}^{F \times 1}$ denotes the frequency steering vector.

Similarly, each interference signal $\mathbf{l}_{j,f,t}$ is modeled as
\begin{equation}
    \mathbf{l}_{j,f,t} = \mathbf{g}_{j,f} s_{j,f,t}, \label{eq:interference_model}
\end{equation}
where $s_{j,f,t}$ is the unknown symbol of the $j$-th interferer at subcarrier $f$ and  OFDM symbol $t$, and $\mathbf{g}_{j,f}$ the channel vector of the $j$-th interferer at subcarrier $f$. The multipath interference channel $\mathbf{g}_{j,f} \in \mathbb{C}^{N \times 1}$ exhibits a geometric structure similar to that of a user channel, i.e.,
\begin{align}
    \mathbf{g}_{j,f} = \sum_{l=1}^{L_j^{\rm int}} \epsilon_{j,l}^{\rm int} \mathbf{a}_{\mathsf{S}}(\theta_{j,l}^{\rm int}, \beta_{j,l}^{\rm int}) e^{-\jmath 2 \pi f \Delta f \tau_{j,l}^{\rm int} },
\end{align}
where $L_j^{\rm int}$ denotes the number of propagation paths of interferer $j$. The parameters $\epsilon_{j,l}^{\rm int}$, $\tau_{j,l}^{\rm int}$, $\theta_{j,l}^{\rm int}$, and $\beta_{j,l}^{\rm int}$ are the complex gain, delay, azimuth, and elevation AoAs for its $l$-th path, respectively. Superscripts $\rm tar$ and $\rm int$ distinguish target-channel and interference-channel geometric parameters throughout the model.
Consequently, the aggregated interference channel matrix of the $j$-th interferer  $\mathbf{G}_{j}={[\mathbf{g}_{j,1}, \ldots, \mathbf{g}_{j, F}]}^{\mathrm{T}} \in \mathbb{C}^{F \times N}$ is given by
\begin{align}
    \mathbf{G}_{j} = \sum_{l=1}^{L_j^{\rm int}} \epsilon_{j,l}^{\rm int} \mathbf{a}_{\mathsf{F}}(\tau_{j,l}^{\rm int}) \mathbf{a}_{\mathsf{S}}{(\theta_{j,l}^{\rm int}, \beta_{j,l}^{\rm int})}^{\mathrm{T}}.
\end{align}
The considered  model is intended for co-channel inter-cell interference generated by OFDM transmitters whose signals propagate to the BS through multipath channels. Arbitrary narrowband, bursty, non-OFDM, or deliberately structured jamming waveforms are outside the main scope, even when they experience multipath propagation.
\subsection{Transmission Frame Structure}
 Each transmission frame is partitioned into two distinct phases: a pilot phase spanning the first $T_{\text{P}}$ OFDM symbols dedicated to channel estimation, and a data transmission phase covering the subsequent $T - T_{\text{P}}$ OFDM symbols.
 
 During the pilot phase, each user transmits a pilot sequence ${\{x_{k,f,t}\}}_{f,t}$ (known by the BS) to facilitate channel estimation. During the data transmission phase, $x_{k,f,t}$ carries the unknown data. The data symbols are assumed to be independent and identically distributed (i.i.d.) and drawn from a complex modulation constellation set $\mathcal{X}_k$ with average power $P_k$, i.e., $\mathbb{E}[|x_k|^2] = P_k$ for $x_k \in \mathcal{X}_k$. Regarding the interference, as the BS lacks prior information concerning the interference symbol, each interference symbol $s_{j,f,t}$ is modeled as an independent random variable with zero mean and variance $\alpha_j$ (i.e., $\mathbb{E}[|s_{j,f,t}|^2] = \alpha_j$) throughout the entire frame\footnote{The variance $\alpha_j$ is unknown at the BS and must be estimated; see~\eqref{eq:em_alpha}.}.
 
 To facilitate further discussion, we aggregate the received pilot signals  into a compact matrix form. First, considering the $t$-th pilot OFDM symbol, we stack the signals across all $F$ subcarriers as
 \begin{align}
   \mathbf{Y}_t = \sum_{k=1}^{K} \mathbf{X}_{k, t}\mathbf{H}_{k}  + \sum_{j=1}^{J} \mathbf{L}_{j, t}  + \mathbf{N}_{t}, \quad 1 \le t \le T_{\text{P}}, 
     \label{eq:matrix_model}
 \end{align}
 where $\mathbf{Y}_t = {[\mathbf{y}_{1,t}, \ldots, \mathbf{y}_{F,t}]}^{\mathsf{T}} \in \mathbb{C}^{F \times N}$ denotes the received signal matrix at time $t$. The diagonal pilot matrix for user $k$ is defined as $\mathbf{X}_{k, t} = \text{diag}\{x_{k,1,t}, \ldots, x_{k, F, t}\} \in \mathbb{C}^{F \times F}$. The frequency-domain channel matrix $\mathbf{H}_{k} \in \mathbb{C}^{F \times N}$, the effective interference channel $\mathbf{L}_{j, t} \in \mathbb{C}^{F \times N}$, and the noise matrix $\mathbf{N}_{t} \in \mathbb{C}^{F \times N}$ are constructed by stacking their respective row vectors ${[\mathbf{h}_{k,1}, \ldots, \mathbf{h}_{k, F}]}^{\mathsf{T}}$, ${[\mathbf{l}_{j,1, t}, \ldots, \mathbf{l}_{j, F, t}]}^{\mathsf{T}}$, and ${[\mathbf{n}_{1,t}, \ldots, \mathbf{n}_{F,t}]}^{\mathsf{T}}$.
 Furthermore, by stacking the observations over the entire pilot duration, the overall system model for the pilot phase can be expressed as
 \begin{align}
     \mathbf{Y} = \sum_{k=1}^{K} \mathbf{X}_{k} \mathbf{H}_{k} + \sum_{j=1}^{J} \mathbf{L}_{j} + \mathbf{N}, 
     \label{eq:full_matrix_model}
 \end{align}
 where the  received signal matrix is $\mathbf{Y} = {[\mathbf{Y}_1^{\mathsf{T}}, \ldots, \mathbf{Y}_{T_{\text{P}}}^{\mathsf{T}}]}^{\mathsf{T}} \in \mathbb{C}^{T_{\text{P}} F \times N}$. Accordingly, the aggregated pilot matrix for user $k$ is formed as $\mathbf{X}_k = {[\mathbf{X}_{k,1}^{\mathsf{T}}, \ldots, \mathbf{X}_{k,T_{\text{P}}}^{\mathsf{T}}]}^{\mathsf{T}} \in \mathbb{C}^{T_{\text{P}} F \times F}$. The interference $\mathbf{L}_j$ and noise $\mathbf{N}$ are similarly defined by stacking $\mathbf{L}_{j,t}$ and $\mathbf{N}_{t}$ for $t=1,\dots,T_{\text{P}}$, respectively.

\subsection{Conventional Receiver Processing}\label{subsec:canonical_rx}

 \begin{figure}[htbp]
     \centering
     \includegraphics[width=0.48\textwidth]{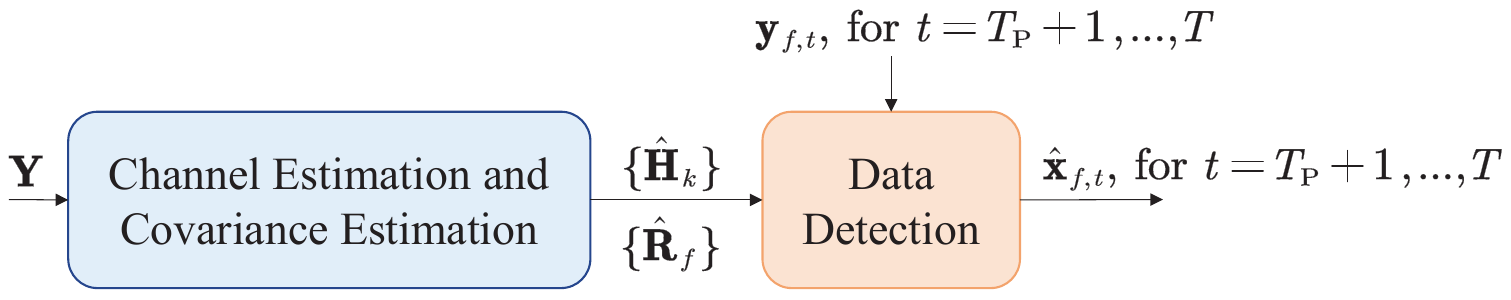}
     \caption{Receiver processing stages consisting of pilot-aided channel estimation and data detection.}\label{fig:receiver_processing}
 \end{figure}
 
 As illustrated in Fig.~\ref{fig:receiver_processing}, we consider a conventional low-complexity two-stage receiver structure consisting of pilot-aided channel estimation and interference-plus-noise covariance
estimation, followed by MMSE-IRC data detection, which is a standard and widely adopted receiver principle in
cellular systems~\cite{tse2005fundamentals,winters1984optimum,3gpp_mmseirc}.

 In the first stage, the receiver performs CSI acquisition based on the received signal matrix $\mathbf{Y}$ during the pilot phase. 
A widely used  baseline is the one-dimensional (1D) frequency-domain LMMSE estimator, which processes each receive antenna independently and ignores spatial channel correlation; more advanced massive MIMO estimators can exploit long-term spatial covariance when reliable covariance estimates are available.
 Let $\mathbf{y}_{r} \in \mathbb{C}^{T_{\text{P}} F \times 1}$ denote  $r$-th column of  $\mathbf{Y}$. The corresponding channel estimate for user $k$ at antenna $r$ is
 \begin{align}
     \hat{\mathbf{h}}_{k,r} &= \mathbf{C}_{k} {\mathbf{X}}_{k}^{\mathrm{H}} {\left( \sum_{i=1}^{K} {\mathbf{X}}_{i} \mathbf{C}_{i}{\mathbf{X}}_{i}^{\mathrm{H}} + \beta\mathbf{I} \right)}^{-1} \mathbf{y}_{r},   
     \label{eq:1d_lmmse}  
 \end{align}
 where $\beta$ represents the  aggregate power of the noise and the interference, and  $\mathbf{C}_{k}$ is the frequency-domain channel covariance matrix determined by the power-delay profile (PDP) $\{\rho_{k,l}, \tau_{k,l}\}$. In practice, the PDP must be estimated and is therefore sensitive to interference contamination.

 To suppress interference during the subsequent data detection stage, the receiver requires  knowledge of the  interference-plus-noise covariance matrix. Ideally, assuming the interference channels $\{\mathbf{g}_{j,f}\}$ are known, the interference-plus-noise covariance matrix at subcarrier $f$ is expressed as 
 \begin{align}
     \mathbf{R}_{f}
     = &\mathbb{E}\!\bigg[
     \Big(\sum_{j=1}^{J}\mathbf{g}_{j,f}s_{j,f,t}+\mathbf{n}_{f,t}\Big)
     \Big(\sum_{j=1}^{J}\mathbf{g}_{j,f}s_{j,f,t}+\mathbf{n}_{f,t}\Big)^{\mathsf H}\nonumber \\
     & \;\;\Big|\{\mathbf{g}_{j,f}\}_{j=1}^{J}
     \bigg] \notag\\
     = &\sum_{j=1}^{J} \alpha_j \mathbf{g}_{j,f} \mathbf{g}_{j,f}^{\mathsf{H}} + \sigma^2 \mathbf{I}, 
     \label{eq:ideal_in_covariance}
 \end{align}
 
 Since the interference channels $\{\mathbf{g}_{j,f}\}$ vary across subcarriers, the optimal $\mathbf{R}_f$ is inherently frequency-selective. However, in practical scenarios where $\{\mathbf{g}_{j,f}\}$ are unavailable, an empirical approximation is typically employed. 
 Specifically,  $ \mathbf{R}_{f}$ is estimated via the sample-average approximation by averaging over a frequency-time resource block (RB)  containing $B$ subcarriers and $T_{\text{P}}$ OFDM symbols~\cite{cho2010mimo}:
 \begin{align}
     \hat{\mathbf{R}}_{f} = \frac{1}{T_{\text{P}} B} \sum_{t=1}^{T_{\text{P}}} \sum_{\nu  \in \mathcal{F}_b} \tilde{\mathbf{y}}_{\nu, t} \tilde{\mathbf{y}}_{\nu, t}^{\mathrm{H}} + \sigma^2\mathbf{I}, 
     \label{eq:subband_scm}
 \end{align}
 where $\mathcal{F}_b$ denotes the set of subcarrier indices assigned to the $b$-th RB, $b = \lceil f/B\rceil$, and $\tilde{\mathbf{y}}_{\nu, t} = \mathbf{y}_{\nu,t} - \sum_{k=1}^{K} \hat{\mathbf{h}}_{k,\nu} x_{k,\nu,t}$ denotes the residual signal obtained by canceling the estimated target signal from the received pilot observations. The RB size $B$ introduces a tradeoff in~\eqref{eq:subband_scm}. 
 A larger $B$ provides more samples and improves estimation robustness, but may blur the frequency-selective structure of the interference covariance. 
 A smaller $B$ better preserves frequency variations, but suffers from insufficient samples for accurate covariance estimation.

 In the second stage, the estimated channel and the interference-plus-noise covariance are employed to perform data detection via the MMSE-IRC detector~\cite{winters1984optimum}. Let $\hat{\mathbf{H}}_{f}\in \mathbb{C}^{N \times K}$ denote the estimated  channel matrix for all $K$ users at subcarrier $f$. The estimate of the  transmitted symbol vector, denoted by $\hat{\mathbf{x}}_{f,t} \in \mathbb{C}^{K \times 1}$, is obtained by
 \begin{align}
     \hat{\mathbf{x}}_{f,t} = \hat{\mathbf{H}}_{f}^{\mathrm{H}} {\left( \hat{\mathbf{H}}_{f} \mathbf{P} \hat{\mathbf{H}}_{f}^{\mathrm{H}} + \hat{\mathbf{R}}_{f} \right)}^{-1} \mathbf{y}_{f,t},
     \label{eq:mmse_irc}
 \end{align}
 $\text{for}\; t = T_{\text{P}}+1, \ldots, T$,  where $\mathbf{P} = \text{diag}\{P_1, \ldots, P_K\}$ represents the diagonal power matrix.

\subsection{Discussion and Motivation}\label{subsec:discussions}
The conventional two-stage receiver in Section~\ref{subsec:canonical_rx} has two limitations under the considered strong-interference setting.

First, for channel estimation, the 1D-LMMSE estimator in~\eqref{eq:1d_lmmse} processes each receive antenna independently and therefore does not exploit the spatial correlation of the propagation environment.
More importantly, its underlying AWGN assumption ignores the  spatially colored nature of interference, which can lead to substantial CSI estimation errors. 
Although many advanced channel estimation methods have been developed to exploit spatial-domain features~\cite{8240714, 11310707, 8400482, 11045666, 10534815, 9685707, 8353153, 10075639}, most of them are derived under interference-free or white-noise assumptions and are thus not well suited to the structured interference considered here. 

Second, for the interference-plus-noise covariance matrix $\hat{\mathbf{R}}_f$, the sample-average approximation in~\eqref{eq:subband_scm} requires averaging over a time-frequency RB. This strategy imposes a locally invariant covariance model over the RB and thus cannot accurately capture the frequency-selective variations of the true interference covariance. 
As a result, the resulting covariance estimate is generally insufficient for fully exploiting the interference rejection capability of the MMSE-IRC detector.

These limitations call for a framework that can jointly estimate the target channels and resolve the interference structure at subcarrier-level resolution.
This observation motivates the joint probabilistic formulation developed in the following sections.

\section{Problem Formulation for Joint Channel Estimation and Interference Suppression}\label{sec:problem_formulation}
We next formulate a probabilistic model that jointly captures the target channels, the interference channels, and the interference symbols. The resulting posterior inference problem is solved by variational inference, and the inferred interference channels and symbols are later used for covariance reconstruction in MMSE-IRC detection.

\subsection{Probability Model}\label{subsec:prob_model}
Instead of treating the interference as Gaussian noise, we preserve its physical bilinear structure by decomposing it into interference channels and stochastic interference symbols. Specifically, the interference matrix $\mathbf{L}_j$ is expressed as
\begin{align}
    \mathbf{L}_{j} = \mathbf{S}_{j} \mathbf{G}_{j},
\end{align}
where $\mathbf{G}_{j}={[\mathbf{g}_{j,1}, \ldots, \mathbf{g}_{j, F}]}^{\mathrm{T}} \in \mathbb{C}^{F \times N}$ denotes the interference channel matrix, capturing the spatial signatures across all subcarriers. The interference symbol matrix is defined as $\mathbf{S}_{j} = {[\mathbf{S}_{j,1}^{\mathrm{T}}, \ldots, \mathbf{S}_{j,T_{\text{P}}}^{\mathrm{T}}]}^{\mathrm{T}} \in \mathbb{C}^{T_{\text{P}} F \times F}$, where the diagonal block for the $t$-th OFDM symbol is given by $ \mathbf{S}_{j, t} = \mathrm{diag}\{s_{j,1,t}, \ldots,s_{j, F, t}\} \in \mathbb{C}^{F \times F}$.
We further define the interference symbol vector $\mathbf{s}_j \in \mathbb{C}^{FT_{\text{P}}}$ by stacking the diagonal entries of $\{\mathbf{S}_{j, t}\}_{t=1}^{T_{\mathsf{P}}}$, i.e., $\mathbf{s}_j = {[s_{j,1,1}, \ldots, s_{j,F,1}, \ldots, s_{j,F, T_{\text{P}}}]}^{\mathrm{T}}$.
The received signal model in~\eqref{eq:full_matrix_model} is reformulated as 
\begin{align}
    \mathbf{Y}  = \sum_{k=1}^{K} \mathbf{X}_{k}\mathbf{H}_{k}  + \sum_{j=1}^{J} \mathbf{S}_{j} \mathbf{G}_{j} + \mathbf{N}. 
    \label{eq:structured_model}
\end{align}

To enable a Bayesian inference framework, we assign probabilistic priors to the unknown variables. The target channel matrix $\mathbf{H}_k$ is modeled hierarchically, where its structure is deterministically defined by $\Gamma_k^{\rm tar} = {\{\beta_{k,l}^{\rm tar}, \tau_{k,l}^{\rm tar}, \theta_{k,l}^{\rm tar}, \varphi_{k,l}^{\rm tar}\}}_{l=1}^{L_k}$ as
\begin{align}
    p(\mathbf{H}_k|\Gamma_k^{\rm tar}) = \!\delta\!\left(\mathbf{H}_k -\sum_{l=1}^{L_k} \beta_{k,l}^{\rm tar}\mathbf{a}_{\mathsf{F}}(\tau_{k,l}^{\rm tar}) {\mathbf{a}_{\mathsf{S}}(\theta_{k,l}^{\rm tar}, \varphi_{k,l}^{\rm tar})}^{\mathsf{T}} \right), \label{eq:likelihood_channel}
\end{align}
Here, $\Gamma_k^{\rm tar}$ collects target-channel parameters and $\Gamma_j^{\rm int}={\{\rho_{j,l}^{\rm int},\tau_{j,l}^{\rm int},\theta_{j,l}^{\rm int},\beta_{j,l}^{\rm int}\}}_{l=1}^{L_j^{\rm int}}$ the analogous interference-channel parameters. Conditioned on these parameters, the channels are deterministically generated by the geometric multipath model.
The target-channel parameters follow $p(\beta_{k,l}^{\rm tar})=\mathcal{CN}(0,\rho_{k,l}^{\rm tar})$, $p(\tau_{k,l}^{\rm tar})=\mathcal{U}(0,\tau_{\max})$, $p(\theta_{k,l}^{\rm tar})=\mathcal{VM}(\mu_{\theta,k,l}^{\rm tar},\kappa_{\theta,k,l}^{\rm tar})$, and $p(\varphi_{k,l}^{\rm tar})=\mathcal{VM}(\mu_{\varphi,k,l}^{\rm tar},\kappa_{\varphi,k,l}^{\rm tar})$.
The prior for $\mathbf{G}_j$ follows the same geometric form as that of $\mathbf{H}_k$ and is omitted for brevity.
Since the receiver has no prior knowledge of the interference symbol, we model $\mathbf{s}_j$ as a zero-mean Gaussian random vector with covariance $\alpha_j \mathbf I$, i.e.,
\begin{equation}
    p(\mathbf{s}_j) = \mathcal{CN}(\mathbf{s}_j; \mathbf{0}, \alpha_{j} \mathbf{I}). \label{eq:prior_interference_symbols}
\end{equation}
This Gaussian model is used as a tractable inference prior for the symbol update and does not require the actual interferer symbols to be Gaussian.
Finally, the additive noise is modeled as 
\begin{equation}
    p(\mathbf{N}) = \prod_{t,f} \mathcal{CN}(\mathbf{n}_{f,t}; \mathbf{0}, \sigma^2 \mathbf{I}).
\end{equation}

Define $\mathcal{H} \triangleq {\{\mathbf{H}_k\}}_k$,   
$\mathcal{G} \triangleq {\{\mathbf{G}_j\}}_j$, 
$\mathcal{S} \triangleq {\{\mathbf{s}_j\}}_j$, 
and the set $\boldsymbol{\Theta} \triangleq \{ \mathcal{H}, \mathcal{G}, \mathcal{S}\}$. Assuming the parameters are statistically independent, the joint prior distribution is factorized as
\begin{align}
    p(\boldsymbol{\Theta}) &  = p(\mathcal{H}) p(\mathcal{G}) p(\mathcal{S})\nonumber \\
    & = \prod_{k=1}^{K} \left( \int p(\mathbf{H}_k|\Gamma_k^{\rm tar}) p(\Gamma_k^{\rm tar}) \, \mathrm{d}\Gamma_k^{\rm tar} \right) \nonumber \\
    & \quad \times \prod_{j=1}^{J} \left( \int p(\mathbf{G}_j|\Gamma_j^{\rm int}) p(\Gamma_j^{\rm int}) \, \mathrm{d} \Gamma_j^{\rm int} \right) \times \prod_{j=1}^{J} p(\mathbf{s}_j),
\end{align}
where $p(\Gamma_k^{\rm tar}) = \prod_{l=1}^{L_k} p(\theta_{k,l}^{\rm tar}) p(\varphi_{k,l}^{\rm tar}) p(\tau_{k,l}^{\rm tar}) p(\beta_{k,l}^{\rm tar})$, and analogously for $p(\Gamma_j^{\rm int})$.
Based on the model in~\eqref{eq:structured_model} and the noise statistics, the likelihood function of the received signal $\mathbf{Y}$, given the system states $\boldsymbol{\Theta}$, is expressed as
\begin{align}
    p(\mathbf{Y} |\boldsymbol{\Theta}) \propto \exp \left( -\frac{1}{\sigma^2} \left\| \mathbf{Y} - \sum_{k=1}^{K} \mathbf{X}_{k}\mathbf{H}_{k} - \sum_{j=1}^{J} \mathbf{S}_{j}\mathbf{G}_{j} \right\|_F^2 \right).
    \label{eq:likelihood}
\end{align}

\subsection{Bayesian Inference}
Our goal is to infer $\boldsymbol{\Theta}=\{\mathcal H,\mathcal G,\mathcal S\}$ from the received pilot observation $\mathbf{Y}$. The exact posterior $p(\boldsymbol{\Theta}|\mathbf{Y})$ is intractable because of the high-dimensional integration and the bilinear coupling among the latent variables. We therefore adopt variational inference and approximate the true posterior by a structured distribution $q(\boldsymbol{\Theta})$, which is optimized in Section~\ref{sec:VI}. The resulting posterior estimates are then used to reconstruct the interference-plus-noise covariance with subcarrier-level resolution for MMSE-IRC detection.

\section{Joint Channel Estimation and Interference Suppression via Score-Based Variational Inference}\label{sec:VI}

    \subsection{Variational Inference Framework}
    According to the system model in~\eqref{eq:full_matrix_model}, the inference task involves the recovery of the target channels $\mathcal{H}$ and the interference components induced by $\{\mathbf{G}_j,\mathbf{S}_j\}_{j=1}^J$.
    A seemingly convenient simplification is to collapse each interference component into the aggregate term $\mathbf{L}_j=\mathbf{S}_j\mathbf{G}_j$ and  apply a naive mean-field approximation~\cite{bishop2006pattern} of the form
    \begin{align}
        q(\mathcal{H}, \{\mathbf{L}_j\} | \mathbf{Y}) 
        = \prod_{k=1}^{K} q(\mathbf{H}_k| \mathbf{Y}) \prod_{j=1}^{J} q(\mathbf{L}_j| \mathbf{Y}),\label{eq:naive_mean_field}
    \end{align}
    However, such a formulation obscures the bilinear structure of the interference. 
    Moreover, treating $\mathbf{L}_j$ as a single latent variable is undesirable, since the randomness of $\mathbf{S}_j$ masks the frequency-domain manifold of the physical channels $\mathbf{G}_j$.
    
    To preserve this structure, we adopt the following structured variational factorization:
    \begin{align}
        q(\boldsymbol{\Theta} | \mathbf{Y})
        = q(\mathcal{H} | \mathbf{Y}) q(\mathcal{G}, \mathcal{S} | \mathbf{Y}),
    \end{align}
    with
    \begin{align}
        q(\boldsymbol{\Theta} | \mathbf{Y})
        =
        \prod_{k=1}^{K} q(\mathbf{H}_k| \mathbf{Y})
        \prod_{j=1}^{J} q(\mathbf{G}_j| \mathbf{Y})
        \prod_{j=1}^{J} q(\mathbf{s}_{j}| \mathbf{Y},\{\mathbf{G}_{j}\}_{j=1}^{J}).
        \label{eq:block_factorization}
    \end{align}
    As a result, the score-based prior is imposed only on the channel variables $\mathbf{H}_k$ and $\mathbf{G}_j$, whereas the interference symbols $\mathbf{s}_j$ are inferred through linear estimation.
    
    The variational objective is to minimize the Kullback-Leibler (KL) divergence:
    \begin{align}
        \hat{q}(\boldsymbol{\Theta}| \mathbf{Y})
        = \arg \min_{q(\boldsymbol{\Theta})}
        \operatorname{KL}(q(\boldsymbol{\Theta}| \mathbf{Y}) \parallel p(\boldsymbol{\Theta} | \mathbf{Y})),
        \label{eq:kl_objective}
    \end{align}
    or equivalently maximize the evidence lower bound~\cite{blei2017variational}:
    \begin{align}
        \mathcal{L}(q(\boldsymbol{\Theta}| \mathbf{Y})) 
        &= \mathbb{E}_q [\ln p(\mathbf{Y} | \boldsymbol{\Theta})]
        - \operatorname{KL}(q(\mathcal{H}| \mathbf{Y}) \parallel p(\mathcal{H})) \nonumber \\
        &\quad
        - \operatorname{KL}(q(\mathcal{G}, \mathcal{S}| \mathbf{Y}) \parallel p(\mathcal{G}, \mathcal{S})).
        \label{eq:elbo}
    \end{align}
    
\begin{figure}[htbp]
        \centering
        \includegraphics[width=0.48\textwidth]{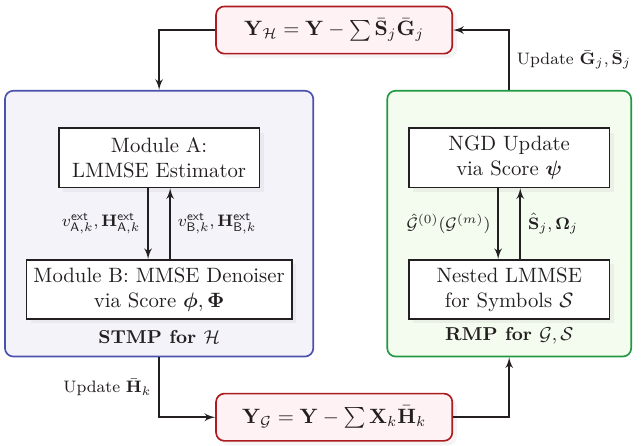}
        \caption{Proposed JCEIS-SVI framework. The algorithm alternates between STMP for target-channel estimation and RMP for interference-channel and symbol inference.}\label{fig:svi_framework}
    \end{figure}

    Due to the bilinear structure in the likelihood, direct optimization of~\eqref{eq:elbo} is intractable. 
    We therefore adopt a block coordinate ascent (BCA) strategy, iteratively updating the target-channel block $\mathcal H$ and the interference block $(\mathcal G,\mathcal S)$:
    \begin{align}
        \ln \hat{q}(\mathcal{H}| \mathbf{Y})
        &\!\propto\!
        \ln p(\mathcal{H})
        + \mathbb{E}_{q(\mathcal{G}, \mathcal{S}| \mathbf{Y})}
        [\ln p(\mathbf{Y} | \mathcal{H}, \mathcal{G}, \mathcal{S})],
        \label{eq:solution_H}\\
        \ln \hat{q}(\mathcal{G}, \mathcal{S}| \mathbf{Y})
        &\!\propto\!
        \ln p(\mathcal{G},\mathcal{S})
        + \mathbb{E}_{q(\mathcal{H}| \mathbf{Y})}
        [\ln p(\mathbf{Y} | \mathcal{H}, \mathcal{G}, \mathcal{S})].
        \label{eq:solution_GS}
    \end{align}
    Equations~\eqref{eq:solution_H} and~\eqref{eq:solution_GS} only specify the general structure of variational factors. 
    To obtain tractable updates, we parameterize the factors as multivariate complex Gaussian distributions. 
    Specifically, we let
    $q(\mathbf{H}_k| \mathbf{Y}) = \mathcal{CN}(\mathbf{H}_k; \bar{\mathbf{H}}_{k}, \mathbf{\Sigma}_{\mathbf{H}_k})$,
    and similarly for $\mathbf{G}_j$ and $\mathbf{s}_{j}$. 
    Consequently, the functional optimization of $q(\cdot)$ is converted into the optimization of its sufficient statistics, namely, the posterior means and covariances. 
    Since the priors of $\mathcal H$ and $\mathcal G$ are highly structured and non-Gaussian, the updates are implemented through score-based modules rather than closed-form Bayesian updates.
    As summarized in Fig.~\ref{fig:svi_framework}, the proposed algorithm alternates between STMP-based channel estimation and RMP-based interference inference.
    
    \subsection{Transformed Formulation for Estimating $\mathcal{H}$}\label{sec:H_estimation}

    Taking the expectation of the log-likelihood term in~\eqref{eq:solution_H}
    with respect to $q(\mathcal{G},\mathcal{S}|\mathbf{Y})$ gives
    \begin{align}
        &\mathbb{E}_{q(\mathcal{G}, \mathcal{S}| \mathbf{Y})}
        \!\left[
        \ln p(\mathbf{Y} | \mathcal{H}, \mathcal{G}, \mathcal{S})
        \right] \nonumber\\
        \propto\;&
        -\frac{1}{\sigma^2}
        \mathbb{E}_{q(\mathcal{G}, \mathcal{S}| \mathbf{Y})}
        \!\left[
        \left\|
        \mathbf{Y}_{\mathcal H}
        - \sum_{k=1}^{K} \mathbf{X}_k \mathbf{H}_k
        - \widetilde{\mathbf L}
        \right\|_\text{F}^2
        \right],
        \label{eq:H_ll_exp}
    \end{align}
    where
    \begin{align}
        \mathbf{Y}_{\mathcal H}
        &\triangleq
        \mathbf Y-
        \mathbb{E}_{q(\mathcal{G}, \mathcal{S}\mid\mathbf{Y})}
        \!\left[
        \sum_{j=1}^{J} \mathbf{S}_j\mathbf{G}_j
        \right], \\
        \widetilde{\mathbf L}
        &\triangleq
        \sum_{j=1}^{J} \mathbf{S}_j\mathbf{G}_j-
        \mathbb{E}_{q(\mathcal{G}, \mathcal{S}\mid\mathbf{Y})}
        \!\left[
        \sum_{j=1}^{J} \mathbf{S}_j\mathbf{G}_j
        \right].
    \end{align}
    Here, $\mathbf{Y}_{\mathcal H}$ is the observation after subtracting the posterior mean of the aggregated interference, and $\widetilde{\mathbf L}$ represents the corresponding residual interference, i.e., the uncertainty in the interference term that is not explained by the current variational estimates of $\mathcal G$ and $\mathcal S$. Let $\mathbf{W}_{\mathcal H}\triangleq \widetilde{\mathbf L}+\mathbf N$ denote the effective disturbance, which consists of the residual interference and the additive noise. To obtain a tractable linear inverse model, we approximate $\mathbf{W}_{\mathcal H}$ as white Gaussian noise, i.e.,
$\mathbf{W}_{\mathcal H}\sim\mathcal{CN}(\mathbf 0,\sigma_{\mathcal H}^2\mathbf I)$.
This white-Gaussian approximation of the residual disturbance is commonly used in iterative Bayesian receivers, where a complicated residual covariance is replaced by a scalar variance or diagonal precision for tractability~\cite{minka2001ep,rangan2011gamp,hagenauer1997turbo}.
Under this approximation, the variational update of $\mathcal H$ is
    \begin{equation}
        \hat{q}(\mathcal{H}\mid\mathbf{Y})
        \propto
        \exp\!\left(
        -\frac{1}{\sigma^2_{\mathcal H}}
        \left\|
        \mathbf{Y}_{\mathcal H}-\sum_{k=1}^{K}\mathbf{X}_k\mathbf{H}_k
        \right\|_\text{F}^2
        \right)
        p(\mathcal{H}).
        \label{eq:H_posterior_form}
    \end{equation}
    This has the standard linear inverse form, corresponding to the auxiliary model
    \begin{equation}
        \mathbf{Y}_{\mathcal H}
        =
        \sum_{k=1}^{K}\mathbf{X}_k\mathbf{H}_k+\mathbf{W}_{\mathcal H}.
        \label{eq:linear_model_H}
    \end{equation}
    To solve this inverse problem with intractable prior $p(\mathcal{H})$, we adopt the STMP framework~\cite{score_CE3, stmp}, which iteratively exchanges information between an LMMSE module (Module A) and a score-based denoising module (Module B).

    \subsection{STMP-Based Estimation of $\mathcal{H}$}
    
    Module A enforces data consistency with respect to the linear model~\eqref{eq:linear_model_H}. 
    In each turbo iteration, the LMMSE posterior mean and variance for $\mathbf{H}_k$ are computed as
    \begin{align}
        \mathbf{H}_{\mathsf{A}, k}^{\mathsf{post}}
        &= \mathbf{H}_{\mathsf{A},k}^{\mathsf{pri}} 
        + 
        v_{\mathsf{A},k}^{\mathsf{pri}} \mathbf{X}_k^{\mathsf{H}}
        \mathbf{\Sigma}_{\mathsf{A}}^{-1}
        \left(
        \mathbf{Y}_{\mathcal{H}} - \sum_{i=1}^{K} \mathbf{X}_i \mathbf{H}_{\mathsf{A},i}^{\mathsf{pri}}
        \right), \label{eq:moduleA_mean} \\
        v_{\mathsf{A},k}^{\mathsf{post}}
        &= v_{\mathsf{A},k}^{\mathsf{pri}}
        - \frac{(v_{\mathsf{A},k}^{\mathsf{pri}})^2}{FN}
        \operatorname{Tr}(\mathbf{X}_k^{\mathsf{H}}\mathbf{\Sigma}_{\mathsf{A}}^{-1} \mathbf{X}_k), \label{eq:moduleA_var}
    \end{align}
    where $\mathbf{\Sigma}_{\mathsf{A}}=
    \sum_{k=1}^{K} v_{\mathsf{A},k}^{\mathsf{pri}} \mathbf{X}_k \mathbf{X}_k^{\mathsf{H}}
    + \sigma^2_{\mathcal{H}} \mathbf{I}$, and the prior mean $\mathbf{H}_{\mathsf{A},k}^{\mathsf{pri}}$ and variance $v_{\mathsf{A},k}^{\mathsf{pri}}$ are provided by Module B. 
    The corresponding extrinsic messages are
    \begin{align}
        v_{\mathsf{A}, k}^{\mathsf{ext}}
        &= \left( \frac{1}{v_{\mathsf{A},k}^{\mathsf{post}}} - \frac{1}{v_{\mathsf{A},k}^{\mathsf{pri}}} \right)^{-1}, \\
        \mathbf{H}_{\mathsf{A},k}^{\mathsf{ext}}
        &= v_{\mathsf{A},k}^{\mathsf{ext}}
        \left(
        \frac{\mathbf{H}_{\mathsf{A},k}^{\mathsf{post}}}{v_{\mathsf{A},k}^{\mathsf{post}}}
        -
        \frac{\mathbf{H}_{\mathsf{A},k}^{\mathsf{pri}}}{v_{\mathsf{A},k}^{\mathsf{pri}}}
        \right).
    \end{align}
    
    Module B treats the extrinsic output of Module A as a pseudo-observation
$\mathbf{H}_{\mathsf{B}, k}^{\mathsf{pri}}=\mathbf{H}_{\mathsf{A},k}^{\mathsf{ext}}
=\mathbf{H}_k + \mathbf{W}_k
$, $ \mathbf{W}_k \sim \mathcal{CN}(\mathbf{0}, v_{\mathsf{B}, k}^{\mathsf{pri}} \mathbf{I})$, with $v_{\mathsf{B}, k}^{\mathsf{pri}} = v_{\mathsf{A}, k}^{\mathsf{ext}}$. 
    Using Tweedie's formula together with the pre-trained score networks $\boldsymbol{\phi}$ and $\boldsymbol{\Phi}$~\cite{stmp}, the MMSE posterior mean and variance are approximated as
    \begin{align}
        \mathbf{H}_{\mathsf{B}, k}^{\mathsf{post}}
        &= \mathbf{H}_{\mathsf{B}, k}^{\mathsf{pri}}
        + v_{\mathsf{B}, k}^{\mathsf{pri}}
        \boldsymbol{\phi}\left(\mathbf{H}_{\mathsf{B}, k}^{\mathsf{pri}}, \sqrt{v_{\mathsf{B}, k}^{\mathsf{pri}}}\right),
        \label{eq:tweedie_update} \\
        v_{\mathsf{B},k}^{\mathsf{post}}
        &= v_{\mathsf{B},k}^{\mathsf{pri}}
        + \frac{(v_{\mathsf{B},k}^{\mathsf{pri}})^2}{FN}
        \operatorname{Tr}\!\left(
        \boldsymbol{\Phi}\left(\mathbf{H}_{\mathsf{B}, k}^{\mathsf{pri}}, \sqrt{v_{\mathsf{B}, k}^{\mathsf{pri}}}\right)
        \right),
        \label{eq:tweedie_var_update}
    \end{align}
    where the pre-trained score networks $\boldsymbol{\phi}(\cdot,\cdot)$ and $\boldsymbol{\Phi}(\cdot,\cdot)$ approximate the first- and second-order score functions of the noisy target-channel distribution, respectively. Specifically, for the pseudo-observation $\tilde{\mathbf H}=\mathbf H+\mathbf W$, where $\mathbf W\sim\mathcal{CN}(\mathbf 0,v\mathbf I)$, the first-order score function is approximated as
   
    \begin{align}
        \nabla_{\tilde{\mathbf H}} \log p_v(\tilde{\mathbf H})
        \approx
        \boldsymbol{\phi}(\tilde{\mathbf H},\sqrt v),
    \end{align}
    where $p_v(\tilde{\mathbf H})$ denotes the distribution of the target channel corrupted by Gaussian noise with variance $v$. Similarly, the second-order score function is approximated by

    \begin{align}
        \nabla_{\tilde{\mathbf H}}^2 \log p_v(\tilde{\mathbf H})
        \approx
        \boldsymbol{\Phi}(\tilde{\mathbf H},\sqrt v).
    \end{align}
    The two networks are trained offline using denoising score matching, following the procedure in~\cite{stmp,sm,lu2022maximum}.
    The extrinsic messages passed back to Module A are
    \begin{align}
        v_{\mathsf{B}, k}^{\mathsf{ext}}
        &= \left( \frac{1}{v_{\mathsf{B},k}^{\mathsf{post}}} - \frac{1}{v_{\mathsf{B},k}^{\mathsf{pri}}} \right)^{-1}, \\
        \mathbf{H}_{\mathsf{B},k}^{\mathsf{ext}}
        &= v_{\mathsf{B},k}^{\mathsf{ext}}
        \left(
        \frac{\mathbf{H}_{\mathsf{B},k}^{\mathsf{post}}}{v_{\mathsf{B},k}^{\mathsf{post}}}
        -
        \frac{\mathbf{H}_{\mathsf{B},k}^{\mathsf{pri}}}{v_{\mathsf{B},k}^{\mathsf{pri}}}
        \right).
        \label{eq:moduleB_extrinsic_mean}
    \end{align}
    After the prescribed STMP iteration(s), we set $\bar{\mathbf H}_k=\mathbf H_{\mathsf B,k}^{\mathsf{post}}$ and $\mathbf\Sigma_{\mathbf H_k}=v_{\mathsf B,k}^{\mathsf{post}}\mathbf I$. The STMP iterations can be terminated when the normalized change of the Module-B posterior output over all target users is smaller than a threshold $\varepsilon_{\rm STMP}$, i.e.,
    \begin{align}
    \delta_{\mathsf B}^{(i)}
    =
    \frac{
    \sum_{k=1}^{K}
    \left\|
    \mathbf H_{\mathsf B,k}^{\mathsf{post},(i)}
    -
    \mathbf H_{\mathsf B,k}^{\mathsf{post},(i-1)}
    \right\|_{\rm F}^2
    }{
    \sum_{k=1}^{K}
    \left\|
    \mathbf H_{\mathsf B,k}^{\mathsf{post},(i-1)}
    \right\|_{\rm F}^2
    }
    < \varepsilon_{\rm STMP}.
    \end{align}

    \subsection{Transformed Formulation for Estimating $\mathcal{G}$ and $\mathcal{S}$}\label{sec:rmp_estimation}
    
    Similarly to~\eqref{eq:H_posterior_form}, by evaluating the expectation in~\eqref{eq:solution_GS} with respect to $q(\mathcal{H}|\mathbf{Y})$, the log-likelihood term associated with the interference block becomes
    \begin{align}
        \mathbb{E}_{q(\mathcal{H}|\mathbf{Y})}
        [\ln p(\mathbf{Y} | \mathcal{H}, \mathcal{S}, \mathcal{G})]
        \propto
        -\frac{1}{\sigma^2_{\mathcal{G}}}
        \left\|
        \mathbf{Y}_{\mathcal{G}}\! -\! \sum_{j=1}^{J} \mathbf{S}_j \mathbf{G}_j
        \right\|_{\text{F}}^2\!,\!
        \label{eq:proxy_likelihood}
    \end{align}
    where
    \begin{align}
        \mathbf{Y}_{\mathcal{G}}
        \triangleq
        \mathbf{Y} - \sum_{k=1}^{K} \mathbf{X}_k \bar{\mathbf{H}}_k
    \end{align}
    is the residual observation after canceling the current estimate of the target signal.
    Therefore, the variational update of the interference block can be expressed as
    \begin{align}
        \hat{q}(\mathcal{G}, \mathcal{S}|\mathbf{Y})
        \propto
        p(\mathbf{Y}_{\mathcal{G}} | \mathcal{G}, \mathcal{S}) \, p(\mathcal{G}) \, p(\mathcal{S}),
        \label{eq:G_posterior_form}
    \end{align}
    with
    \begin{align}
        p(\mathbf{Y}_{\mathcal{G}} | \mathcal{G}, \mathcal{S})
        \propto
        \exp\left(
        -\frac{1}{\sigma^2_{\mathcal{G}}}
        \left\|
        \mathbf{Y}_{\mathcal{G}} - \sum_{j=1}^{J} \mathbf{S}_j \mathbf{G}_j
        \right\|_{\text{F}}^2
        \right).
        \label{eq:YG_likelihood_form}
    \end{align}
    
    Unlike the $\mathcal{H}$-subproblem, which reduces to a linear inverse problem, the inference of $(\mathcal{G},\mathcal{S})$ is a bilinear matrix factorization problem. 
    This structure yields a highly nonconvex posterior landscape and introduces the well-known permutation and phase ambiguities~\cite{8063440, 8764368}. 
    To exploit the conditional structure of the interference block, we organize the inference according to the chain rule,
    \begin{align}
        q(\mathcal{G}, \mathcal{S}|\mathbf{Y})
        =
        q(\mathcal{G} | \mathbf{Y}_{\mathcal{G}})
        q(\mathcal{S} | \mathbf{Y}_{\mathcal{G}}, \mathcal{G}),
        \label{eq:posterior_decomposition}
    \end{align}
    where the channel variables $\mathcal{G}$ are updated by a score-based reverse-diffusion process, while the conditional symbol posterior is handled analytically through an LMMSE step.

    This leads to an RMP module. Specifically, RMP regularizes the interference channels using a learned score prior, while a nested LMMSE step updates the interference symbols conditioned on the current channel estimate. The resulting symbol estimates are then fed back to construct the likelihood score used in the update of the interference channels.
    
    \subsection{RMP-Based Estimation of $\mathcal{G}$ and $\mathcal{S}$}
    
    \subsubsection{Reverse-Diffusion Variational Formulation}
    The RMP framework \cite{rmp} approximates the posterior mean of $p(\mathcal{G} | \mathbf{Y}_{\mathcal{G}})$ by reversing a discrete-time diffusion trajectory. Let $\{\mathcal{G}^{(m)}\}_{m=0}^{M}$ denote the diffusion trajectory, where $\mathcal{G}^{(m)} \triangleq \{\mathbf{G}_1^{(m)}, \dots, \mathbf{G}_{J}^{(m)}\}$ represents the collective interference channels at the $m$-th noise level.

    The forward process incrementally injects Gaussian noise into the clean channel $\mathcal{G}^{(0)} = \mathcal{G}$ until the structured manifold is transformed into a tractable  distribution (e.g., pure noise) at $\mathcal{G}^{(M)}$. For each interference $j$, the forward transition kernel is defined as
    \begin{align}
    p(\mathbf{G}_j^{(m+1)}|\mathbf{G}_j^{(m)})\! = \!\mathcal{CN}(\mathbf{G}_j^{(m+1)}; \!\mathbf{G}_j^{(m)}, \!(v_{m+1}\! -\! v_{m})\mathbf{I}),
    \end{align}
    where  $v_M > v_{M-1} > \cdots > v_0=0$ denote pre-defined noise variances.  The joint posterior distribution is expressed as 
    \begin{align}
        p(\mathcal{G}^{(0:M)} | \mathbf{Y}_{\mathcal{G}}) \!= \!p(\mathcal{G}^{(M)}| \mathbf{Y}_{\mathcal{G}}) \!\!\prod_{m=0}^{M-1}\! p(\mathcal{G}^{(m)} | \mathcal{G}^{(m+1)}, \mathbf{Y}_{\mathcal{G}}).
    \end{align}
    
    We approximate  $p(\mathcal{G}^{(0:M)} | \mathbf{Y}_{\mathcal{G}})$ with a structured variational distribution $q(\mathcal{G}^{(0:M)} | \mathbf{Y}_{\mathcal{G}})$, which is factorized as
    \begin{align}
        & q(\mathcal{G}^{(0:M)} | \mathbf{Y}_{\mathcal{G}}) \nonumber \\
        &   = q_M(\mathcal{G}^{(M)} | \mathbf{Y}_{\mathcal{G}}) \prod_{m=0}^{M-1} q_m(\mathcal{G}^{(m)} | \mathcal{G}^{(m+1)}, \mathbf{Y}_{\mathcal{G}}).
    \end{align}
    Following the RMP framework in~\cite{rmp}, each reverse transition step is modeled as a Gaussian distribution given by
    \begin{align}
        q_m(\mathcal{G}^{(m)} | \mathcal{G}^{(m+1)}, \mathbf{Y}_{\mathcal{G}}) = \prod_{j=1}^{J} \mathcal{CN}(\mathbf{G}_j^{(m)}; \bar{\mathbf{G}}_j^{(m)}, \mathbf{\Lambda}_{m}^{-1}). \label{eq:q_m}
    \end{align}
    where $\bar{\mathbf{G}}_j^{(m)}$ is the variational mean to be estimated, and the precision is approximated by $\mathbf{\Lambda}_m = (v_{m+1}-v_{m})^{-1}\mathbf{I} $~\cite{rmp}.
    
    Our objective is to infer the posterior means $\{\bar{\mathbf{G}}_j^{(m)}\}_{j=1}^{J}$ by minimizing the KL divergence over the entire diffusion trajectory expressed as 
    \begin{align}
        &\hat{q}(\mathcal{G}^{(0:M)} | \mathbf{Y}_{\mathcal{G}}) \nonumber \\
        =& \arg\!\!\!\min_{q(\mathcal{G}^{(0:M)} | \mathbf{Y}_{\mathcal{G}}) } \!\!\!\operatorname{KL}\left( q(\mathcal{G}^{(0:M)} | \mathbf{Y}_{\mathcal{G}}) \parallel p(\mathcal{G}^{(0:M)}  | \mathbf{Y}_{\mathcal{G}}) \right).
    \end{align}
    As established in~\cite[Proposition 2]{rmp}, this global objective can be decoupled into a sequence of step-wise sub-problems solved in reverse order ($m = M-1, \dots, 0$), given by
    \begin{align}
          &\min_{q_m} \operatorname{KL}\!\!\left( \!q_m(\mathcal{G}^{(m)} | \mathcal{G}^{(m+1)}, \!\mathbf{Y}_{\mathcal{G}})  \!\!\parallel\!  p(\mathcal{G}^{(m)} | \mathcal{G}^{(m+1)}, \!\mathbf{Y}_{\mathcal{G}}) \!\right)\!.
        \label{eq:rmp_kl}
    \end{align}
    This recursive decoupling enables an iterative refinement of the interference channels, as detailed in the following subsections.
    \subsubsection{Natural-Gradient Update of the Interference Channels}
    
    The variational means $\{\bar{\mathbf{G}}_j^{(m)}\}_{j=1}^{J}$ are updated by natural gradient descent (NGD)~\cite{martens2020new}. 
    At diffusion step $m$, the update takes the form~\eqref{eq:rmp_update}, where $\eta_m$ is the step size and $\zeta$ is the number of Monte Carlo samples. 
    \begin{figure*}[t]
        \centering
    \begin{align}
        \bar{\mathbf{G}}_{j}^{(m)}
        \leftarrow
        \bar{\mathbf{G}}_{j}^{(m)}
        +
        \eta_m \mathbf{\Lambda}_m^{-1}
        \frac{1}{\zeta}\sum_{l=1}^{\zeta}
        \nabla_{\mathbf{G}_{j}^{(m)}}\log p(\mathcal{G}^{(m)} | \mathcal{G}^{(m+1)}, \mathbf{Y}_{\mathcal{G}})
        \Big|_{\mathcal{G}^{(m)} = \mathcal{G}^{(m,l)} \sim q_m},
        \quad j = 1,\ldots, J.
        \label{eq:rmp_update}
    \end{align}
    \hrulefill
    \end{figure*}
    By Bayes' rule, the reverse-posterior score in~\eqref{eq:rmp_update} can be decomposed as
    \begin{align}
         &\nabla_{\mathbf{G}_{j}^{(m)}} \log p(\mathcal{G}^{(m)} | \mathcal{G}^{(m+1)}, \mathbf{Y}_{\mathcal{G}})
        =
        \nabla_{\mathbf{G}_{j}^{(m)}} \log p(\mathbf{G}_{j}^{(m)}) \nonumber \\
        &+
        \nabla_{\mathbf{G}_{j}^{(m)}} \log p(\mathbf{G}_{j}^{(m+1)} | \mathbf{G}_{j}^{(m)})
        +
        \nabla_{\mathbf{G}_{j}^{(m)}} \log p(\mathbf{Y}_{\mathcal{G}} | \mathcal{G}^{(m)}).
        \label{eq:gradient_decomposition}
    \end{align}
    These three terms play distinct roles:
    
    \emph{Prior score:} the first term regularizes the update toward the learned distribution of physical interference channels. 
    It is parameterized by the pre-trained score network $\boldsymbol{\psi}(\mathbf{G}_{j}^{(m)}, \sqrt{v_m})$.
    
    \emph{Transition score:} the second term enforces consistency with the forward diffusion kernel. 
    For the Gaussian transition, it admits the closed form
    $(v_{m+1} - v_m)^{-1}(\mathbf{G}_{j}^{(m+1)} - \mathbf{G}_{j}^{(m)})$~\cite{rmp}.
    
    \emph{Likelihood score:} the third term injects the measurement information from $\mathbf{Y}_{\mathcal{G}}$ into the reverse trajectory. 
    This term is the key distinction from standard unconditional diffusion sampling and requires a dedicated treatment because the likelihood depends on the latent symbols $\mathcal{S}$.
    
    \subsubsection{Likelihood score via LMMSE symbol inference}
    
    The likelihood in~\eqref{eq:YG_likelihood_form} is defined with respect to the clean channel variables $\mathcal{G}^{(0)}$, whereas the likelihood score in~\eqref{eq:rmp_update} is conditioned on the noisy diffusion state $\mathcal{G}^{(m)}$. 
    Following~\cite{rmp, dps}, we first estimate the clean channels from the noisy state using Tweedie's formula:
    \begin{align}
        \hat{\mathbf{G}}_{j}^{(0)}(\mathbf{G}_j^{(m)})
        =
        \mathbf{G}_{j}^{(m)} + v_m \boldsymbol{\psi}(\mathbf{G}_{j}^{(m)}, \sqrt{v_m}).
        \label{eq:tweedie_est}
    \end{align}
    The likelihood score is then approximated through these clean estimates $\hat{\mathcal{G}}^{(0)} = \{\hat{\mathbf{G}}_{j}^{(0)}\}$ as
    \begin{align}
        &\;\nabla_{\mathbf{G}_{j}^{(m)}} \log p(\mathbf{Y}_{\mathcal{G}} | \mathcal{G}^{(m)}) \nonumber \\
         \approx & \;
        \nabla_{\mathbf{G}_{j}^{(m)}}\log p(\mathbf{Y}_{\mathcal{G}} | \hat{\mathcal{G}}^{(0)}), \nonumber \\
        =& \;
        \mathbb{E}_{\mathcal{S} | \mathbf{Y}_{\mathcal{G}}, \hat{\mathcal{G}}^{(0)}}
        \left[
        \nabla_{\mathbf{G}_{j}^{(m)}}
        \log p(\mathbf{Y}_{\mathcal{G}} | \mathcal{S}, \hat{\mathcal{G}}^{(0)})
        \right] \nonumber \\
        =&
        - \frac{1}{\sigma^2_{\mathcal{G}}}
        \nabla_{\mathbf{G}_{j}^{(m)}}
        \left\|
        \mathbf{Y}_{\mathcal{G}} - \sum_{i} \hat{\mathbf{S}}_i \hat{\mathbf{G}}_{i}^{(0)}
        \right\|_F^2 \nonumber \\
        &
        - \frac{1}{\sigma^2_{\mathcal{G}}}
        \nabla_{\mathbf{G}_{j}^{(m)}}
        \left(
        \sum_{u,i}
        \operatorname{Tr}\left(
        (\hat{\mathbf{G}}_{u}^{(0)})^{\mathsf{H}}
        \mathbf{\Omega}_{ui}
        \hat{\mathbf{G}}_{i}^{(0)}
        \right)
        \right),
        \label{eq:chain_rule}
    \end{align}
    where 
    $\hat{\mathbf{S}}_i$ is the posterior mean of $\mathbf{S}_i$, and $\mathbf{\Omega}_{ui}$ is the posterior error cross-covariance between $\mathbf{S}_u$ and $\mathbf{S}_i$. 
    The derivation of the last step of~\eqref{eq:chain_rule} is given in Appendix~\ref{appendix:likelihood_score}.
    Because the symbol prior is Gaussian, these posterior statistics can be obtained in closed form by an LMMSE step. 
    For the $(t,f)$-th time-frequency grid, the local observation model is
    \begin{align}
        \mathbf{z}_{t,f} = \mathbf{A}_{f} \mathbf{s}_{t,f} + \mathbf{w}_{ t,f},
    \end{align}
    where $\mathbf{z}_{t,f} \in \mathbb{C}^{N\times 1}$ is the transpose of the $((t-1)F+f)$-th row of $\mathbf{Y}_{\mathcal{G}}$,
    $\mathbf{s}_{t,f} = [s_{1,t,f}, \dots, s_{J,t,f}]^{\mathsf T}$ is the vector of symbols transmitted by the $J$ interferers at time $t$ and frequency $f$,
   $\mathbf{A}_{f} =
    [(\hat{\mathbf{G}}_{1}^{(0)})_f^{\mathsf T}, \dots, (\hat{\mathbf{G}}_{J}^{(0)})_f^{\mathsf T}]
    \in \mathbb{C}^{N \times J}$ is the effective channel matrix at frequency $f$ formed by stacking the $f$-th rows of the clean-channel estimates,
    and $\mathbf{w}_{t,f} \sim \mathcal{CN}(\mathbf{0}, \sigma^2_{\mathcal{G}} \mathbf{I})$ is the effective noise.
    The LMMSE posterior statistics are
    \begin{subequations}
    \begin{align}
        \hat{\mathbf{s}}_{t,f}
        &=
        \left(
        \mathbf{C}_s^{-1}
        + \frac{1}{\sigma^2_{\mathcal{G}} } \mathbf{A}_f^{\mathsf{H}} \mathbf{A}_f
        \right)^{-1}
        \frac{1}{\sigma^2_{\mathcal{G}} } \mathbf{A}_f^{\mathsf{H}} \mathbf{z}_{t,f},
        \label{eq:lmmse_mean}\\
        \mathbf{V}_{t,f}
        &=
        \left(
        \mathbf{C}_s^{-1}
        + \frac{1}{\sigma^2_{\mathcal{G}} } \mathbf{A}_f^{\mathsf{H}} \mathbf{A}_f
        \right)^{-1},
        \label{eq:lmmse_cov}
    \end{align}
    \end{subequations}
    where $\mathbf{C}_s=\operatorname{diag}(\alpha_1, \dots, \alpha_{J})$ denotes the prior covariance of the interference symbols. 
    The global quantities $\hat{\mathbf{S}}_i$ and $\mathbf{\Omega}_{ui}$ are then assembled from the local posterior means and covariances over all $(t,f)$.
    
    The above approximation replaces $p(\mathcal{S}|\mathbf{Y}_{\mathcal{G}},\mathcal{G})$ with $p(\mathcal{S}| \mathbf{Y}_{\mathcal{G}},\hat{\mathcal{G}}^{(0)})$ throughout the RMP iterations, thereby creating a synergistic inference loop. 
    As the reverse diffusion process progressively refines the channel manifold, the symbol inference becomes more accurate; in turn, the refined symbol posterior sharpens the likelihood score and steers the diffusion trajectory toward the true interference manifold. 
    The final estimates of the interference channels and symbols are given by
    $\bar{\mathbf{G}}_{j}= \bar{\mathbf{G}}_{j}^{(0)}$ and
    $\bar{\mathbf{S}}_{j}= \mathbb{E}[\mathbf{S}_j | \mathbf{Y}_{\mathcal{G}}, \{\bar{\mathbf{G}}_{j}\}]$,
    respectively.
    
    \subsection{Overall Algorithm}\label{sec:overall_algorithm}
    The overall JCEIS-SVI procedure is summarized in Algorithm~\ref{alg:joint_algo}. 
    The STMP module (Lines 7--10) updates the target-channel posterior, while the RMP module (Lines 12--22) updates the interference channels and the interference symbols. 
    We adopt a nested alternating schedule between the STMP and RMP modules. 
    Specifically, for each reverse diffusion step $m$, the RMP module performs $D$ inner iterations (i.e., NGD updates). 
    At each inner iteration, the interference-channel variables $\mathcal{G}$ and the interference-symbol variables $\mathcal{S}$ are updated using their current posterior estimates.  Every $I_{\text{freq}}$ inner iterations, the current interference estimates are then used to construct the effective observation   $\mathbf{Y}_{\mathcal{H}} \approx \mathbf{Y} - \sum_{j=1}^{J}\bar{\mathbf{S}}_j\bar{\mathbf{G}}_j$, based on which one STMP update is performed for the target channels $\mathcal{H}$.
    The effective noise variances are updated using the posterior uncertainty returned by the corresponding modules. Specifically,
    \begin{align}
        \hat{\sigma}_{\mathcal H}^2
        =
        \frac{1}{T_{\mathrm P}FN}
        \left\|
        \mathbf{Y}
        - \sum_{k=1}^{K} \mathbf{X}_k \bar{\mathbf{H}}_k
        - \sum_{j=1}^{J} \bar{\mathbf{S}}_j \bar{\mathbf{G}}_j
        \right\|_\text{F}^2, \label{eq:sigma_H_update}
    \end{align}
    and
    \begin{align}
        \hat{\sigma}_{\mathcal G}^2
        =
        \sigma^2
        +
        \frac{1}{T_{\mathrm P}FN}
        \sum_{k=1}^{K} v_{\mathsf{B},k}^{\mathsf{post}}\|\mathbf X_k \|^2_{\text{F}}.
        \label{eq:sigma_G_update}
    \end{align}
    For each interference $j$, the interference symbol power $\alpha_j$ is updated via an EM step~\cite{bishop2006pattern}:
    \begin{align}
        \hat{\alpha}_j
        =
        \frac{1}{F T_{\text{P}}}
        \sum_{f=1}^{F} \sum_{t=1}^{T_{\text{P}}}
        \left( |\hat{s}_{j,f,t}|^2 + [\mathbf{V}_{t,f}]_{j,j} \right).
        \label{eq:em_alpha}
    \end{align}

A strict convergence proof for the overall algorithm is difficult
because the framework involves several approximations,
including partial alternating updates, learned score functions,
and the likelihood-score approximation in the reverse process.
Nevertheless, the annealed diffusion schedule provides
effective regularization in practice, and the resulting algorithm
exhibits stable empirical behavior, as shown in Section~\ref{sec:numerical}.
    
Finally, we emphasize  that the estimated interference channels  and symbols may deviate from their ground-truth values because of the inherent bilinear ambiguities of matrix factorization.  
Specifically, any pair $(\mathbf{S}_j\mathbf{Q}, \mathbf{Q}^{-1} \mathbf{G}_j )$ yields the same product as the original pair $(\mathbf{S}_j, \mathbf{G}_j)$, provided that $\mathbf{Q} = \mathbf{\Xi}\mathbf{\Pi}$ is a scaled permutation matrix, where $\mathbf{\Xi}$   and $\mathbf{\Pi}$ denote a diagonal phase-shift matrix and a permutation matrix, respectively~\cite{8063440, 8764368}. Hence,  $\mathbf{G}_j$ and $\mathbf{S}_j$ are identifiable only up to  phase rotations and ordering, whereas their product $\mathbf{S}_j \mathbf{G}_j$ and consequently the aggregate interference $\sum_{j} \mathbf{S}_j \mathbf{G}_j$, remains invariant. Accordingly, the proposed receiver does not aim to identify the physical channel and symbol sequence of each interferer individually. The estimated factors provide a virtual decomposition of the aggregate interference, and their individual identities are auxiliary to the inference procedure. For target-channel estimation, only the reconstructed aggregate interference \(\sum_{j}\hat{\mathbf{S}}_j\hat{\mathbf{G}}_j\) is cancelled from the received signal. Hence, phase and permutation ambiguities of individual components do not affect the target-channel estimate as long as the aggregate dominant interference is accurately reconstructed. For the purpose of IRC, an exact reconstruction of $\mathbf{G}_j$ and $\mathbf{S}_j$ is in fact not necessary. The MMSE-IRC receiver depends  primarily on the interference subspace characterized by its covariance matrix. Phase rotations and permutations preserve this subspace and
the aggregate interference covariance. Our framework focuses on robustly capturing this subspace to construct a precise subcarrier-level interference covariance matrix, as further elaborated in Section~\ref{sec:IRC}.

    \begin{algorithm}[htbp]
        \caption{Proposed JCEIS-SVI Algorithm}\label{alg:joint_algo}
        \begin{algorithmic}[1]
        \REQUIRE~$\mathbf{Y}$, $\{\mathbf{X}_k\}$,$\{\boldsymbol{\phi}, \boldsymbol{\Phi}, \boldsymbol{\psi}\}$, ${\{v_m\}}_{m=0}^M$, $I_{\text{freq}}$, $J$.
        \ENSURE~$\{\bar{\mathbf{H}}_k\}$, $\{\bar{\mathbf{G}}_j\}$, $\{\bar{\mathbf{S}}_j\}$.
        \STATE~\textbf{Initialize:} 
               $\mathbf{H}_{\mathsf{A}, k}^{\mathsf{pri}} = \mathbf{0}$, $v_{\mathsf{A}, k}^{\mathsf{pri}}=1$ ($\forall k$);
               $\alpha_j = 1$,  $\bar{\mathbf{S}}_j = \mathbf{0}$ ($\forall j$);
                $\bar{\mathbf{G}}_j^{(M)} \sim \mathcal{CN}(\mathbf{0}, v_M \mathbf{I})$ and $\bar{\mathbf{G}}_j = \mathbf{0}$ ($\forall j$).
        \FOR{$m = M-1$ \TO~$0$}
            \FOR{$d=1$ \TO~$D$}
                \IF{$d \pmod{I_{\text{freq}}} == 0$}
                    \STATE Compute clean interference-channel estimates $\bar{\mathbf G}_j \leftarrow \hat{\mathbf G}_j^{(0)}(\bar{\mathbf G}_j^{(m)})$, $\forall j$.
                    \STATE~Update interference symbol posterior $\bar{\mathbf S}_j \leftarrow \mathbb E[\mathbf S_j\mid \mathbf Y_{\mathcal G},\{\bar{\mathbf G}_j\}]$, $\forall j$.

                    \STATE~Cancel interference: $\mathbf{Y}_{\mathcal{H}} \leftarrow \mathbf{Y} - \sum_{j=1}^{J}\bar{\mathbf{S}}_j\bar{\mathbf{G}}_j$.
                    \STATE~Update effective noise variance $\sigma_{\mathcal{H}}^2$ via Eq.~\eqref{eq:sigma_H_update}.
                    \STATE~Run one iteration of STMP (Eqs.~\eqref{eq:moduleA_mean}-\eqref{eq:moduleB_extrinsic_mean}) using effective observation $\mathbf{Y}_{\mathcal{H}}$.
                    \STATE~Update channel estimates: $\bar{\mathbf{H}}_k \leftarrow \mathbf{H}_{\mathsf{B},k}^{\mathsf{post}}$, $\forall k$.
                \ENDIF
                
                \STATE~Cancel target signals: $\mathbf{Y}_{\mathcal{G}} \leftarrow \mathbf{Y} - \sum_{k=1}^{K}\mathbf{X}_k \bar{\mathbf{H}}_k$.
                \STATE~Update effective noise variance $\sigma_{\mathcal{G}}^2$ via Eq.~\eqref{eq:sigma_G_update}.
                \STATE~Sample $\mathcal{G}^{(m)}$ from $q_{m}(\mathcal{G}^{(m)} |\mathcal{G}^{(m+1)}, \mathbf{Y}_{\mathcal{G}})$.
                \STATE~Compute Tweedie estimate $\hat{\mathbf{G}}^{(0)}_j$ via Eq.~\eqref{eq:tweedie_est}, $\forall j$.
                \STATE~Compute LMMSE symbol mean $\hat{\mathbf{s}}_{f,t}$ and variance $\mathbf{V}_{f,t}$ via Eqs.~\eqref{eq:lmmse_mean} and~\eqref{eq:lmmse_cov}, $\forall f, t$.
                \STATE~Compute likelihood score via Eqs.~\eqref{eq:tweedie_est}-\eqref{eq:chain_rule}, $\forall j$.
                \STATE~Update variational mean $\bar{\mathbf{G}}_j^{(m)}$ via Eq.~\eqref{eq:rmp_update}, $\forall j$.
                \STATE~Update symbol power $\alpha_j$ via Eq.~\eqref{eq:em_alpha}, $\forall j$.
            \ENDFOR
            \STATE~Update interference channel estimates for step $m$: $\bar{\mathbf{G}}_j \leftarrow \hat{\mathbf{G}}_j^{(0)}(\bar{\mathbf{G}}_j^{(m)})$, $\forall j$.
            \STATE~Update interference symbol estimates for step $m$: $\bar{\mathbf{S}}_j \leftarrow \mathbb{E}[\mathbf{S}_j | \mathbf{Y}_{\mathcal{G}}, \{\bar{\mathbf{G}}_j\}]$, $\forall j$.
        \ENDFOR
        \STATE~\textbf{Return} $\{\bar{\mathbf{H}}_k\}$, $\{\bar{\mathbf{G}}_j\}$, $\{\bar{\mathbf{S}}_j\}$.
        \end{algorithmic}
    \end{algorithm}

\section{Interference-Subspace-Tracking-Based Data Detection}\label{sec:IRC}
After estimating the interference channels $\mathcal{G}$ and symbols $\mathcal{S}$, the receiver proceeds to the data detection stage. 
As discussed in Section~\ref{sec:overall_algorithm}, although the individual interference factors may not be uniquely identifiable because of bilinear ambiguities, the resulting interference subspace remains invariant. 
Motivated by this observation, we develop an interference-subspace-tracking-based IRC scheme. 
By reconstructing the interference and exploiting its spectral correlation, the proposed method forms a refined interference-plus-noise covariance matrix for MMSE-IRC detection.

\subsection{Interference Reconstruction and Subspace Extraction}
Unlike the sample-average method in~\eqref{eq:subband_scm}, which relies on noisy residual observations after target-signal cancellation, the proposed  approach reconstructs the interference directly from the estimated $\mathcal{G}$ and $\mathcal{S}$. 
Let $\hat{\mathbf{G}}_{f} \in \mathbb{C}^{N \times J}$ and $\hat{\mathbf{s}}_{f,t} \in \mathbb{C}^{J \times 1}$ denote the estimated interference channel matrix and the corresponding interference symbol vector at subcarrier $f$ and OFDM symbol $t$, respectively. 
The aggregate interference vector at subcarrier $f$ and OFDM symbol $t$ is reconstructed as
\begin{equation}
    \hat{\mathbf{l}}_{f,t} = \hat{\mathbf{G}}_{f}\hat{\mathbf{s}}_{f,t},
    \qquad t=1,\dots,T_{\text{P}},\; f=1,\dots,F.
    \label{eq:interference_reconstruction}
\end{equation}
To characterize the spatial directions occupied by the interference, we construct the orthogonal projection matrix onto the estimated interference subspace as
\begin{equation}
    \mathbf{\Upsilon}_{f}
    =
    \hat{\mathbf{G}}_{f}
    \left(\hat{\mathbf{G}}_{f}^{\mathsf H}\hat{\mathbf{G}}_{f}\right)^{\dagger}
    \hat{\mathbf{G}}_{f}^{\mathsf H},
    \label{eq:proj_matrix}
\end{equation}
where $(\cdot)^{\dagger}$ denotes the Moore-Penrose pseudoinverse.

\subsection{Covariance Matrix Reconstruction}
To exploit the spectral correlation of the interference power profile, we apply a sliding window $\mathcal{W}_f=\{k \mid |k-f| \le \lfloor |\mathcal{W}_f|/2 \rfloor\}$.
The smoothed average interference power is computed as
\begin{equation}
    \bar{p}_{f}
    =
    \frac{1}{|\mathcal{W}_f|T_{\text{P}}J}
    \sum_{i \in \mathcal{W}_f}\sum_{t=1}^{T_{\text{P}}}
    \|\hat{\mathbf{l}}_{i,t}\|^2.
    \label{eq:freq_smoothing_power}
\end{equation}
The interference-plus-noise covariance matrix is then reconstructed as
\begin{equation}
    \hat{\mathbf{R}}_{f}
    =
    \bar{p}_{f}\mathbf{\Upsilon}_{f}
    +
    \sigma^2\mathbf{I}.
    \label{eq:cov_final}
\end{equation}
Here, $\mathbf{\Upsilon}_{f}$ captures the estimated interference subspace, while $\bar{p}_{f}$ provides frequency-smoothed power level for that subspace. 
This scalar projection model is a low-parameter approximation.
The resulting covariance matrix is finally used in the MMSE-IRC detector~\eqref{eq:mmse_irc} to recover the target signals.

\section{Numerical Results}\label{sec:numerical}
In this section, we evaluate the proposed algorithm through numerical simulations.
\subsection{Simulation Setup}

The primary simulation parameters are summarized in Table~\ref{tab:sim_params}. 
We consider a BS equipped with a $4\times 32$ cross-polarized uniform planar array (UPA), yielding $N=256$ antenna elements. 
A comb pilot pattern is adopted, where pilot symbols are inserted every $D_{\text{p}}=K$ subcarriers for each target user. 
The full-band frequency-domain channels are reconstructed by linear interpolation from the estimates at the pilot positions.

The channels are generated using the QuaDRiGa toolbox~\cite{Quadriga2023} according to the 3GPP UMa-NLoS model~\cite{3gpp.38.901} at a carrier frequency of $6.7$~GHz. 
The BS antenna height is fixed at $25$~m, while both the target users and the interference sources are placed at a height of $1.5$~m. 
The target users are randomly dropped within a  distance range of $[50,200]$~m from the BS, whereas the interference sources are distributed within $[200,300]$~m. 
The root-mean-square (RMS) delay spreads are set to $100$~ns for the target channels and $200$~ns for the interference channels. 
Each channel realization contains $20$ clusters, and each cluster consists of $20$ sub-paths. 
The spatial-temporal parameters, including AOAs, delays, and path gains, are randomly generated to provide diverse channel realizations.

The score networks $\boldsymbol{\phi}$, $\boldsymbol{\Phi}$, and $\boldsymbol{\psi}$ adopt the NCSN++ backbone~\cite{song2021scorebased}. 
These networks are trained using denoising score matching~\cite{stmp,sm,lu2022maximum} with the Adam optimizer, a learning rate of $10^{-4}$, a batch size of $128$, and $50$ training epochs. 
Training stability is further improved by using an exponential moving average with a decay rate of $0.999$. Two separate channel databases are generated for training: the target-channel database is used for $\boldsymbol{\phi}$ and $\boldsymbol{\Phi}$, while the interference-channel database is used for $\boldsymbol{\psi}$. The two databases follow the corresponding target-user and interference-source geometry described above, and each database contains $16{,}000$ training samples and $500$ test samples.
To stabilize score-network training, each channel matrix in the training set is normalized to maintain a consistent energy scale across different propagation conditions.
At runtime, the input of the score networks is evaluated on the same normalized channel scale as in training, without explicitly rescaling the channel: the power variation of the target channel is absorbed by the symbol power $P_k$, and that of the interference channel by the corresponding power factor $\alpha_j$.

For the proposed algorithm, the STMP module uses one iteration with an update interval of $I_{\text{freq}}=4$. 
In the RMP module, the number of reverse diffusion steps is set to $M=30$, and each noise level is associated with $D=15$ NGD updates. 
The NGD step size is fixed at $\eta=\eta_m=0.3$, with $\zeta=1$ sample per step~\cite{rmp}. 
The noise schedule $\{v_m\}$ follows a geometric progression between $v_{\min}=10^{-3}$ and $v_{\max}=10$~\cite{song2021scorebased}. 
For the MMSE-IRC detector, the subband size and the sliding-window size are set to $B=12$ and $|\mathcal{W}_f|=12$, respectively.
All simulations are conducted on a Linux server equipped with Intel Core i9-13900K CPUs and NVIDIA RTX 3090 GPUs.

Both target and interference data symbols use 16-QAM, whereas their pilot symbols use QPSK.
For  data detection evaluation, each target user transmits 88 information bits per frame,
LDPC-coded at an eﬀective rate of 88/192 and mapped to 16-QAM over 48 sub-carriers
in a single data OFDM symbol; soft LLRs are computed by approximate demodulation,
and belief-propagation decoding with up to 50 iterations is applied.

\begin{table}[htbp]
\caption{Simulation Parameters}\label{tab:sim_params}
\centering
\footnotesize
\begin{tabular}{ll}
\toprule
\textbf{Parameter} & \textbf{Value}\\
\midrule
BS antennas ($N$) & $4 \times 32 \times 2 = 256$ (Cross-pol UPA) \\
Target users ($K$) & 4 \\
Interference sources ($J$) & 4 \\
Subcarriers ($F$) & 48 \\
Subcarrier spacing ($\Delta f$) & 30 kHz \\
Carrier frequency & 6.7 GHz \\
OFDM symbols ($T$) & 2 \\
Pilot symbols ($T_{\mathsf{P}}$) & 1 \\
Channel coding & 1/2-rate LDPC~\cite{3gpp.38.212} \\
Modulation & 16-QAM (Data), QPSK (Pilot) \\
Channel model & 3GPP TR UMa-NLoS~\cite{3gpp.38.901} \\
\bottomrule
\end{tabular}
\end{table}

\subsection{Evaluation Metrics and Benchmark Schemes}
We evaluate the CE performance in terms of NMSE, and the data detection performance in terms of BLER. 
For clarity, the signal-to-interference ratio (SIR) is defined as the ratio of  the target-signal power and the interference power, i.e.,
\begin{align}
    \mathrm{SIR}
    =
    \frac{\sum_{k=1}^{K} \mathbb{E}\left[\|\mathbf{X}_k \mathbf{H}_k\|^2_{\text{F}}\right]}
    {\sum_{j=1}^{J} \mathbb{E}\left[\|\mathbf{S}_j \mathbf{G}_j\|^2_{\text{F}}\right]}.
\end{align}
Similarly, the interference-to-noise ratio (INR) is defined as
\begin{align}
    \mathrm{INR}
    =
    \frac{\sum_{j=1}^{J} \mathbb{E}[\|\mathbf{S}_j \mathbf{G}_j\|^2_{\text{F}}]}
    {NT_{\mathrm{P}}F\sigma^2}.
\end{align}

\subsubsection{Channel-Estimation Benchmarks}

The proposed \textbf{JCEIS-SVI} is compared with the following benchmark schemes.
\begin{itemize}
    \item \textbf{1D-LMMSE}: The conventional frequency-domain LMMSE estimator defined in~\eqref{eq:1d_lmmse}.
    
    \item \textbf{Genie-aided LMMSE (G-LMMSE)}: An idealized LMMSE estimator with perfect knowledge of the target-channel spatio-frequency covariance and the per-subcarrier interference-plus-noise covariance given by \eqref{eq:ideal_in_covariance}. This scheme serves as a genie-aided benchmark with ideal second-order statistics.
    
    \item \textbf{VALSE}~\cite{valse}: A gridless variational Bayesian framework for super-resolution line spectral estimation, which infers channel parameters in the continuous-parameter domain.
    
    \item \textbf{ARA}~\cite{ARA}: A deep-learning-based baseline that employs an attentive residual autoencoder to exploit spatio-frequency correlations for channel denoising.
     
    \item \textbf{VBI-DSAE}~\cite{DSAE}: A cascaded baseline that first performs LMMSE filtering using a delay-domain-averaged spatial interference covariance estimate, and then refines the channel estimate using an autoencoder.
    
\end{itemize}

\subsubsection{Ablation Variants}
To assess the contribution of individual components in the proposed framework, we  further consider the following ablation variants:
\begin{itemize}
    \item \textbf{STMP}: The RMP module is removed, and only the STMP module is retained for target-channel estimation.
    
    \item \textbf{STMP + Aggregate-RMP}: The RMP module is retained, but infers the aggregated interference $\{\mathbf{L}_j\}$ based on the variational factorization~\eqref{eq:naive_mean_field}. This variant is used to evaluate the benefit of explicitly modeling the interference symbols in the RMP module.
    
    \item \textbf{STMP + Genie-RMP}: The likelihood score is computed with respect to the interference channels $\{\mathbf{G}_j\}$ under perfect knowledge of the interference symbols $\mathcal{S}$. This variant serves as a genie-aided benchmark for the RMP module and isolates the impact of symbol uncertainty.
    
    \item \textbf{STMP (Interference-Free)}: The STMP module is applied to the interference-free observation. This variant serves as a genie-aided benchmark for the STMP module and isolates the impact of interference.
\end{itemize}

\subsubsection{Interference-Covariance Estimation Benchmarks}

For MMSE-IRC detection, we compare the proposed projection-based covariance reconstruction (\textbf{Proj-}$\mathbf{R}$) with the following alternatives:
\begin{itemize}
    \item \textbf{Ave-}$\mathbf{R}$: The interference-plus-noise covariance estimated by the sample-average approximation in~\eqref{eq:subband_scm}.
    
    \item \textbf{Ideal-}$\mathbf{R}$: The exact interference-plus-noise covariance matrix given by~\eqref{eq:ideal_in_covariance}, which serves as a theoretical  bound.

\end{itemize}

\subsection{Ablation Study}

\begin{figure}
    \centering
    \includegraphics[width=0.8\linewidth]{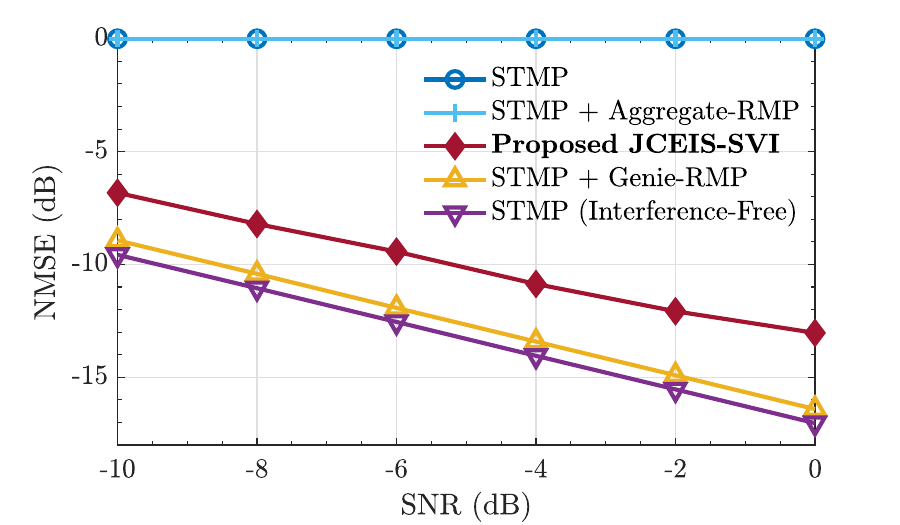}
    \caption{NMSE ablation study illustrating the contributions of the STMP and RMP modules at $\mathrm{INR}=10$~dB.}\label{fig:ablation_nmse}
\end{figure}

We first evaluate STMP and RMP with $\mathrm{INR}=10$~dB. Fig.~\ref{fig:ablation_nmse} shows that STMP and STMP + Aggregate-RMP provide limited gains because dynamic symbols obscure the channel structure in $\mathbf L_j=\mathbf S_j\mathbf G_j$. Explicitly inferring channels and symbols instead gives the proposed RMP more than $9$~dB improvement at $\mathrm{SNR}=-6$~dB. The gap between STMP + Genie-RMP and STMP (Interference-Free) is below $0.5$~dB.

\begin{figure}
    \centering
    \includegraphics[width=0.8\linewidth]{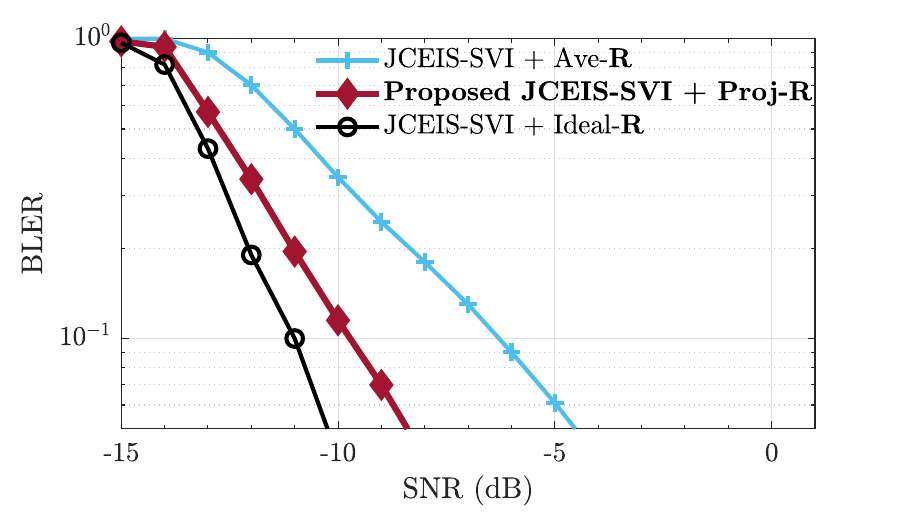}
    \caption{BLER comparison of different interference covariance reconstruction schemes at $\mathrm{INR}=10$~dB.}\label{fig:ablation_bler}
\end{figure}

For data detection, Fig.~\ref{fig:ablation_bler} shows that Proj-$\mathbf R$ outperforms Ave-$\mathbf R$ by about $3$~dB at a BLER of $10^{-1}$ and remains close to Ideal-$\mathbf R$, confirming accurate interference-subspace reconstruction.

\subsection{CE Performance}

\begin{figure}[t]
    \centering
    \includegraphics[width=0.99\linewidth]{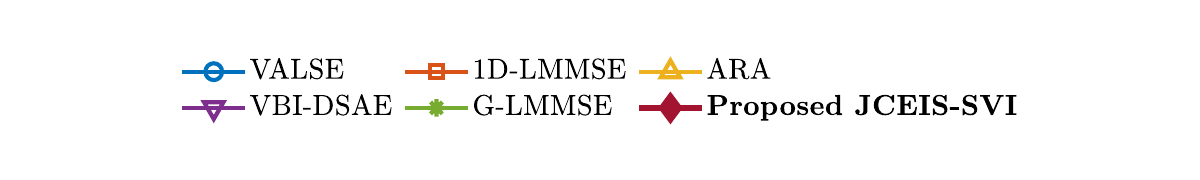} \\
    \vspace{-0.55cm} 
    
    \subfloat[]{
        \centering
        \includegraphics[width=0.30\linewidth]{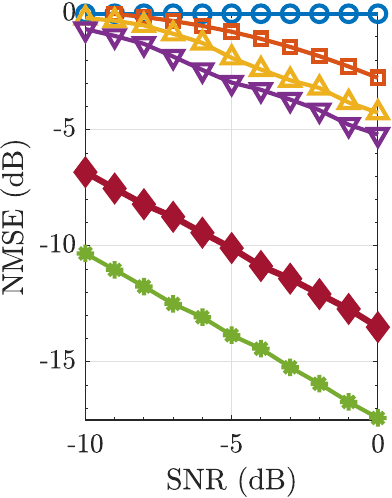}
        \label{fig:ce_snr}
        }
    \subfloat[]{
        \centering
        \includegraphics[width=0.30\linewidth]{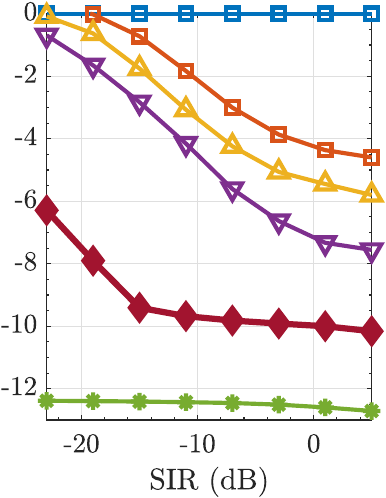}\label{fig:ce_sir}
        }
    \subfloat[]{
            \centering
            \includegraphics[width=0.30\linewidth]{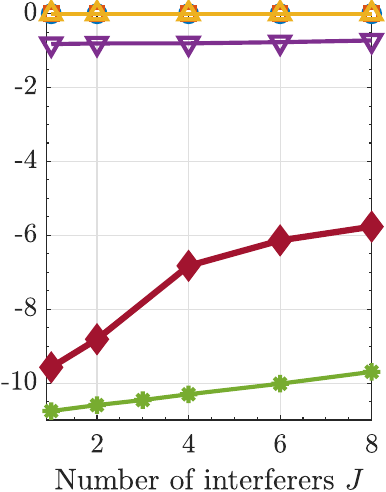}
\label{fig:ce_j}
        }
    
        \caption{CE performance under different interference conditions. (a) NMSE versus SNR ($\mathrm{INR}=10$~dB, $J=4$). (b) NMSE versus SIR ($\mathrm{SNR}=-7$~dB, $J=4$). (c) NMSE versus $J$ ($\mathrm{SNR}=-10$~dB, $\mathrm{INR}=10$~dB).}
    \label{fig:nmse_performance}
\end{figure}

Fig.~\ref{fig:nmse_performance}(a) compares the CE NMSE versus SNR with the INR fixed at $10$~dB. The corrected G-LMMSE curve, which uses genie-aided target and per-subcarrier interference statistics, provides the expected lower bound. Among the practical schemes,
the proposed JCEIS-SVI achieves the lowest NMSE over
the entire SNR range, closely approaching the G-LMMSE
bound (e.g., within about 4 dB at SNR = $-6$ dB) and
outperforming the remaining baselines by a large margin (e.g.,
about 12 dB over VBI-DSAE). This advantage stems from the score-based prior and explicit interference inference, which capture structured interference beyond conventional second-order spatio-domain modeling.

The NMSE-versus-SIR behavior is examined in Fig.~\ref{fig:nmse_performance}(b) with the SNR fixed at $-7$~dB. Among the practical schemes, JCEIS-SVI maintains the best estimation accuracy even in
strongly interference-limited regimes. In particular, to achieve a target NMSE of $-6$~dB, it requires about $15$~dB lower SIR than VBI-DSAE. As the SIR increases, all methods gradually approach their noise-limited floors, while the genie-aided G-LMMSE remains the performance bound because it uses perfect second-order statistics.

Regarding the number of interferers, Fig.~\ref{fig:nmse_performance}(c) shows the NMSE at the SNR   $=-10$~dB and the INR  $=10$~dB. While other
benchmarks vary only mildly with $J$, both G-LMMSE and
JCEIS-SVI degrade as $J$ increases because the interference occupies
a richer spatial subspace and induces stronger coupling
with the target channel. Nevertheless, JCEIS-SVI remains the
best among the practical schemes and stays close to the G-LMMSE
lower bound for all tested values of $J$.

\subsection{Data Detection Performance}

\begin{figure}[H] 
    \centering

    \includegraphics[width=0.99\linewidth]{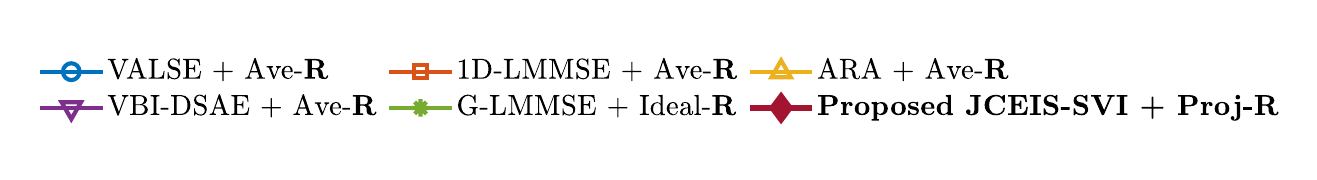} \\
    \vspace{-0.55cm}

    \subfloat[]{
        \centering
        \includegraphics[width=0.30\linewidth]{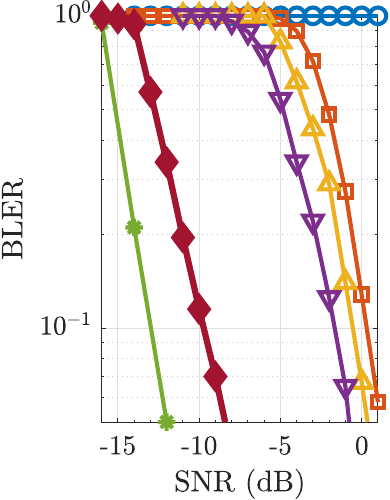}
        \label{fig:bler_snr}
        }
    \subfloat[]{        
        \centering
        \includegraphics[width=0.30\linewidth]{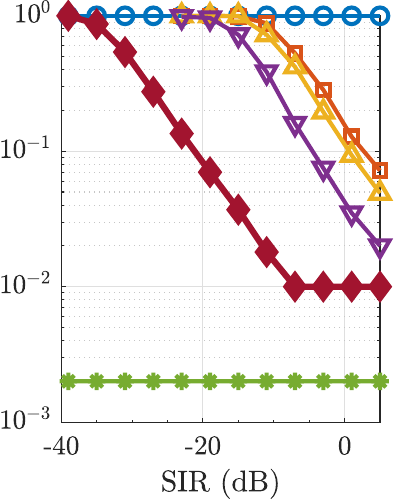}
        \label{fig:bler_sir}
        }
    \subfloat[]{
        \centering
        \includegraphics[width=0.30\linewidth]{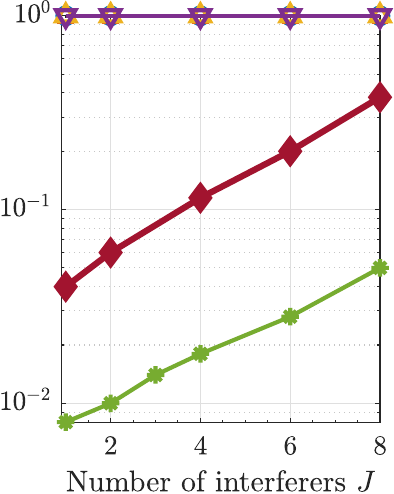}
        \label{fig:bler_j}
        }

    \caption{Data detection performance under different interference conditions. (a) BLER versus SNR ($\mathrm{INR}=10$~dB, $J=4$). (b) BLER versus SIR ($\mathrm{SNR}=-7$~dB, $J=4$). (c) BLER versus $J$ ($\mathrm{SNR}=-10$~dB, $\mathrm{INR}=10$~dB).}
    \label{fig:bler_performance}
\end{figure}

Fig.~\ref{fig:bler_performance}(a) shows that the proposed receiver achieves the
lowest BLER among practical schemes, closely approaching
the G-LMMSE bound with Ideal-$\mathbf{R}$.  At a target BLER of $10^{-1}$, JCEIS-SVI provides a gain of $7$~dB over VBI-DSAE with Ave-$\mathbf{R}$. This result indicates that the improved channel estimation accuracy and interference characterization achieved by the proposed JCEIS-SVI translate directly into more reliable data detection.

Fig.~\ref{fig:bler_performance}(b) evaluates the robustness against increasingly strong interference. The proposed JCEIS-SVI with Proj-$\mathbf{R}$ remains effective even at $\mathrm{SIR}=-20$~dB and outperforms the schemes using Ave-$\mathbf{R}$. For example, at a target BLER of $10^{-1}$, it provides more than $15$~dB SIR gain over VBI-DSAE with Ave-$\mathbf{R}$. Moreover, the performance gap between JCEIS-SVI with Proj-$\mathbf{R}$ and G-LMMSE with Ideal-$\mathbf{R}$ gradually narrows as the SIR increases, which is consistent with the NMSE convergence trend in Fig.~\ref{fig:nmse_performance}(b).

Fig.~\ref{fig:bler_performance}(c) shows that the BLER increases with the number of interferers $J$, as the interference occupies a larger spatial subspace. JCEIS-SVI outperforms the practical baselines, demonstrating robust operation in higher-dimensional interference scenarios.

\subsection{Interferer-Number Mismatch}
\begin{figure}[t]
        \centering
        \includegraphics[width=0.7\linewidth]{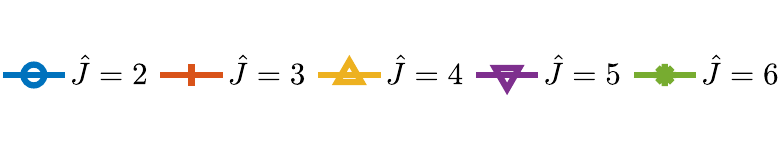} \\
        \vspace{-0.55cm}

        \subfloat[]{
            \centering
            \includegraphics[width=0.45\linewidth]{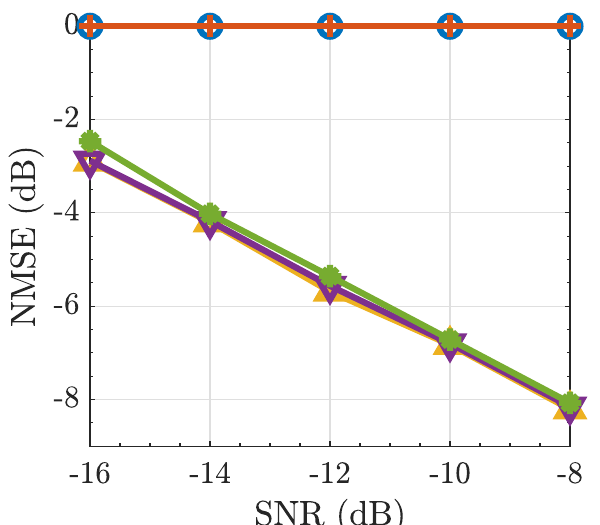}\label{fig:J_nmse}
            }
        \subfloat[]{
            \centering
            \includegraphics[width=0.45\linewidth]{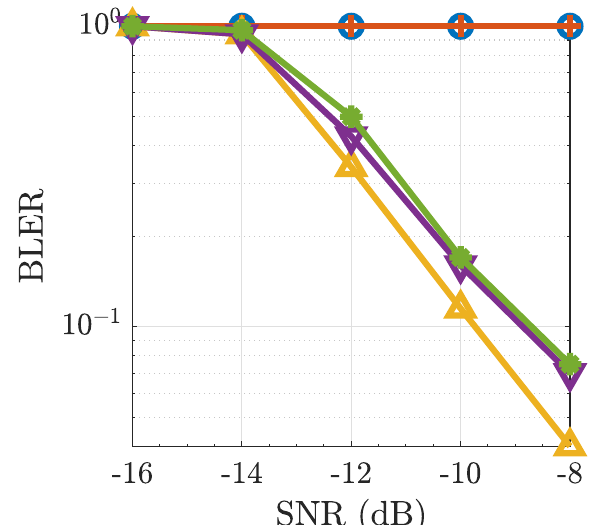}\label{fig:J_bler}
            }
        \caption{NMSE and BLER versus SNR under different estimated numbers of interferers, with the true number fixed at $J=4$ and $\mathrm{INR}=10$~dB. (a) NMSE. (b) BLER.}
        \label{fig:J_performance}
    \end{figure}

To assess sensitivity to mismatch in the estimated interferer number, we fix the ground truth at $J=4$. As illustrated in Fig.~\ref{fig:J_performance}, overestimation ($\hat J=5$ or $6$) leaves NMSE nearly unchanged but slightly degrades detection through over-whitening, whereas underestimation ($\hat J=2$ or $3$) effectively prevents target-channel recovery: NMSE is around $0$~dB and BLER approaches $1$. Thus, slight overestimation is preferable to underestimation.

\subsection{Robustness to Interference Delay- and Angular-Spread Mismatch}
We next evaluate the robustness of the pre-trained interference score prior to static distribution mismatch in the interference delay spread and interference angular spread. The score prior is trained under an RMS delay spread (DS) of 200~ns and an azimuth spread of arrival (ASA) of $30^\circ$. Without retraining, the runtime DS is varied over 30, 100, 200, 300, and 400~ns, while the runtime ASA is varied over $10^\circ$, $20^\circ$, $30^\circ$, $45^\circ$, and $60^\circ$. The remaining configuration is unchanged, with $\mathrm{INR}=10$~dB and $J=4$.

\begin{figure}[!t]
\centering
    \subfloat[DS: NMSE.]{\includegraphics[width=0.48\linewidth]{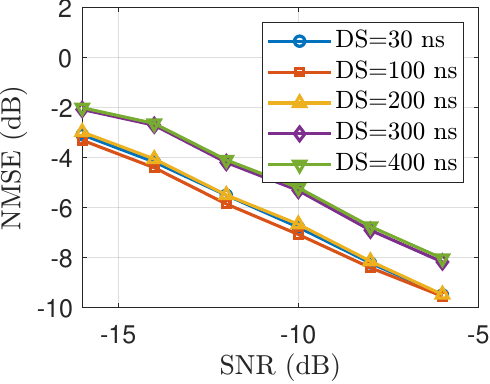}\label{fig:ds_mismatch_nmse}}
    \hfill
    \subfloat[DS: BLER.]{\includegraphics[width=0.48\linewidth]{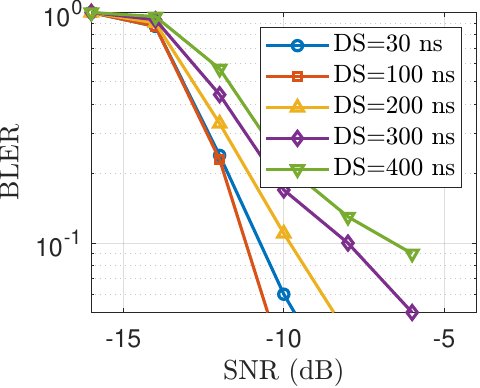}\label{fig:ds_mismatch_bler}}\\[-1mm]
    \subfloat[ASA: NMSE.]{\includegraphics[width=0.48\linewidth]{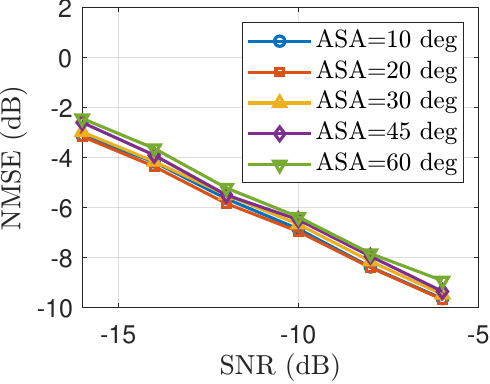}\label{fig:asa_mismatch_nmse}}
    \hfill
    \subfloat[ASA: BLER.]{\includegraphics[width=0.48\linewidth]{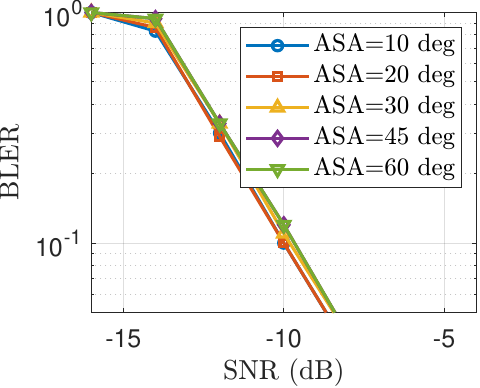}\label{fig:asa_mismatch_bler}}
    \caption{Target-channel NMSE and BLER under interference-channel distribution mismatch ($\mathrm{INR}=10$~dB, $J=4$): (a)--(b) RMS delay spread (DS), and (c)--(d) azimuth spread of arrival (ASA).}
    \label{fig:ds_asa_mismatch}
\end{figure}

Figs.~\ref{fig:ds_asa_mismatch}(a) and~\ref{fig:ds_asa_mismatch}(b) show the target-channel NMSE and receiver BLER. The DS dependence is non-monotonic because a reduced DS has two opposing effects: it increases frequency-domain correlation across subcarriers, which facilitates inference, but also increases mismatch from the training distribution. Consequently, reducing the DS from 200 to 100~ns slightly improves NMSE, whereas the larger mismatch at 30~ns offsets part of this correlation gain. In contrast, increasing the DS to 300 or 400~ns both weakens the frequency-domain correlation and increases the mismatch, resulting in a gradual performance degradation.

The receiver is also robust to interference angular-spread mismatch. 
Over the tested ASA range, the target-channel NMSE variation is below approximately 1~dB.
These results demonstrate that the proposed receiver remains effective under the tested static mismatches in interference delay spread and interference angular spread, without retraining the networks.

\subsection{Sensitivity to RMP Hyperparameters}

\begin{figure*}[!t]
    \centering
    \subfloat[$M$: NMSE.]{\includegraphics[width=0.24\textwidth]{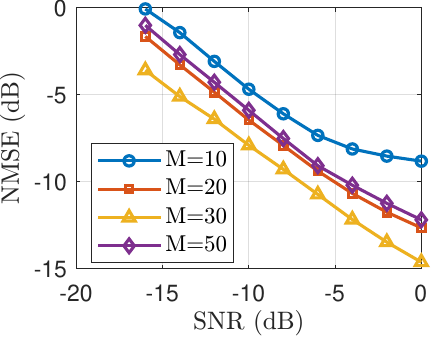}}
    \hfill
    \subfloat[$D$: NMSE.]{\includegraphics[width=0.24\textwidth]{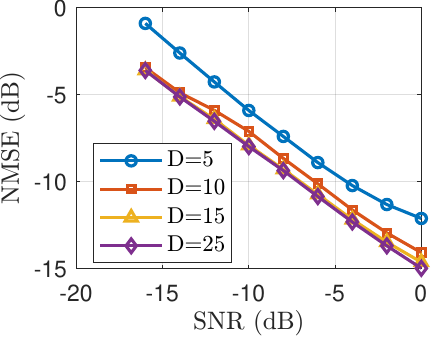}}
    \hfill
    \subfloat[$\eta$: NMSE.]{\includegraphics[width=0.24\textwidth]{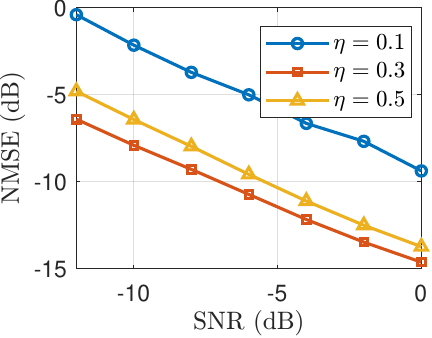}}
    \hfill
    \subfloat[Noise-schedule: NMSE.]{\includegraphics[width=0.24\textwidth]{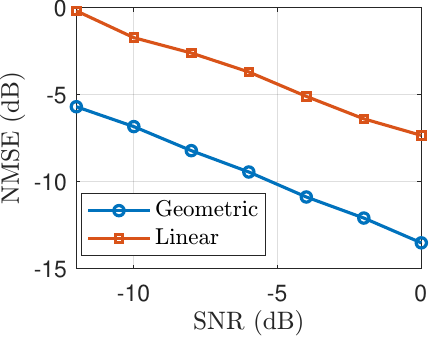}}\\[-1mm]
    \subfloat[$M$: BLER.]{\includegraphics[width=0.24\textwidth]{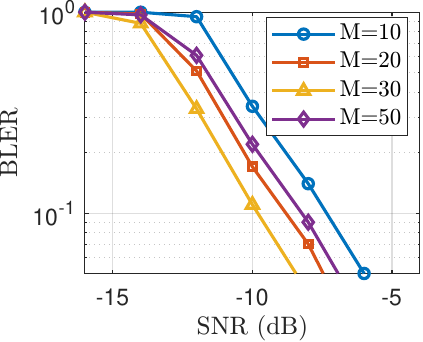}}
    \hfill
    \subfloat[$D$: BLER.]{\includegraphics[width=0.24\textwidth]{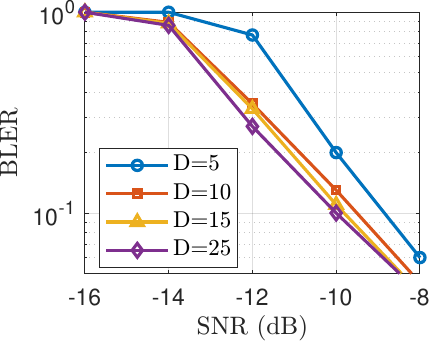}}
    \hfill
    \subfloat[$\eta$: BLER.]{\includegraphics[width=0.24\textwidth]{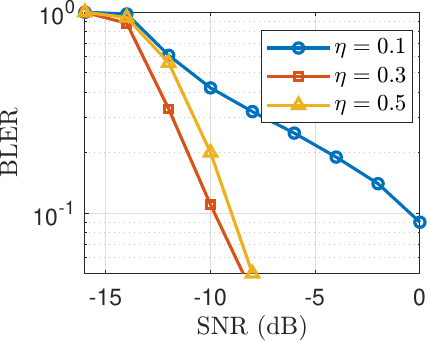}}
    \hfill
    \subfloat[Noise-schedule: BLER.]{\includegraphics[width=0.24\textwidth]{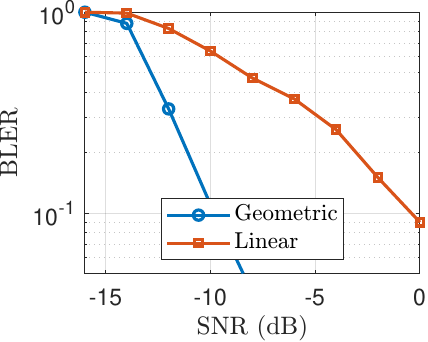}}
    \caption{RMP hyperparameter sensitivity: target-channel NMSE and BLER versus SNR}
    \label{fig:cost_quality_ablation}
\end{figure*}

The online cost of RMP is mainly determined by the number of reverse-diffusion steps $M$ and the number of NGD inner updates $D$. For each inferred interference-channel sample, the number of score-network evaluations is proportional to $MD$. In Fig.~\ref{fig:cost_quality_ablation}, we vary $M$ over $\{10,20,30,50\}$, $D$ over $\{5,10,15,25\}$, and the NGD step size $\eta$ over $\{0.1,0.3,0.5\}$. We also compare geometric and linear noise schedules. The default setting uses $M=30$, $D=15$, and $\eta=0.3$. All results use $\mathrm{INR}=10$~dB and $J=4$.

Reducing $M$ from $30$ to $20$ decreases the score-network cost by one third, but causes a moderate performance loss. At $\mathrm{SNR}=0$~dB, the target-channel NMSE changes from $-14.61$ to $-12.64$~dB, while the BLER at $\mathrm{SNR}=-10$~dB increases from $0.11$ to $0.17$. Setting $M=10$ severely degrades the reconstruction. Increasing $M$ to $50$ provides no improvement and slightly worsens NMSE while increasing the computational cost. Thus, too few reverse steps leave the channel insufficiently refined, whereas excessively many steps do not improve the approximation.

The method is less sensitive to $D$ over the range from $10$ to $25$. Reducing $D$ from $15$ to $10$ lowers the cost by one third with a negligible NMSE loss: at $\mathrm{SNR}=0$~dB, the NMSE changes only from $-14.61$ to $-14.06$~dB. In contrast, $D=5$ leaves the NGD refinement insufficient, while $D=25$ yields only marginal improvement. The default $D=15$ is therefore a robust choice, and $D=10$ provides a lower-complexity alternative.

For the NGD step size, $\eta=0.1$ is too small to provide sufficient refinement within the available inner iterations, whereas $\eta=0.5$ causes mild overshoot. The default $\eta=0.3$ achieves the best overall NMSE--BLER tradeoff.

We further compare the geometric variance-exploding noise schedule used by the proposed receiver~\cite{song2021scorebased},
\[
v_m=v_{\min}\left(\frac{v_{\max}}{v_{\min}}\right)^{m/(M-1)},
\]
with the linear schedule~\cite{karras2022elucidating},
\[
v_m=v_{\min}+(v_{\max}-v_{\min})\frac{m}{M-1}.
\]
The geometric schedule allocates more reverse-diffusion resolution to the low-noise regime, where fine channel structure is recovered. The linear schedule consequently degrades both NMSE and BLER. At $\mathrm{SNR}=-12$~dB, the NMSE changes from $-5.68$ to $-0.18$~dB; at $\mathrm{SNR}=-10$~dB, the BLER increases from $0.11$ to $0.63$.

Table~\ref{tab:md_runtime} reports the measured RMP wall-clock runtime per realization on the same GPU platform. Each column contains paired $M$ and $D$ settings with the same product $MD$, and their runtimes are nearly identical. This confirms that the RMP runtime grows approximately linearly with $MD$, increasing from $26.7$~s at $MD=150$ to $133.6$~s at $MD=750$.

\begin{table}[t]
\caption{RMP Wall-Clock Cost for $M$--$D$ Sweeps}\label{tab:md_runtime}
\centering\scriptsize\renewcommand{\arraystretch}{0.76}
\begin{tabular}{lcccc}
\toprule
Paired setting & 1 & 2 & 3 & 4\\
\midrule
$M$ ($D=15$) & 10 & 20 & 30 & 50\\
$D$ ($M=30$) & 5 & 10 & 15 & 25\\
$MD$ & 150 & 300 & 450 & 750\\
RMP runtime (s) & 26.7 & 53.4 & 80.2 & 133.6\\
\bottomrule
\end{tabular}
\end{table}

\subsection{RMP Convergence Trajectories}
\begin{figure}[!t]
    \centering
    \includegraphics[width=0.55\linewidth]{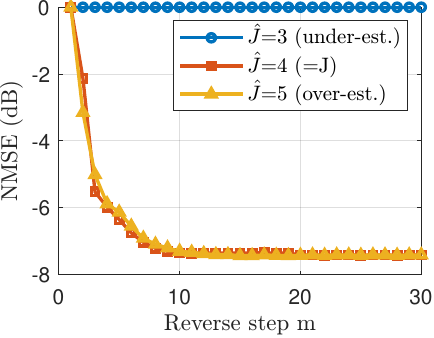}
    \caption{Target-channel NMSE along the reverse-diffusion trajectory for the true, underestimated, and overestimated interferer numbers.}
    \label{fig:rmp_trajectory}
\end{figure}

 Fig.~\ref{fig:rmp_trajectory} records the NMSE after each reverse-diffusion step at $\mathrm{SNR}=-10$~dB and $\mathrm{INR}=10$~dB. The true number of interference streams is $J=4$, while the assumed value is set to $\hat{J}=3$, $4$, or $5$ to examine underestimation, matched model order, and mild overestimation.

For the matched setting $\hat{J}=4$, the NMSE decreases rapidly during the early reverse-diffusion steps and then stabilizes at approximately $-7.3$~dB. The trajectory for $\hat{J}=5$ is nearly identical, showing that mild overestimation preserves the dominant interference subspace and has little effect on the target-channel estimate. In contrast, when $\hat{J}=3$, the assumed interference subspace is insufficient to represent the dominant interference. The target-channel NMSE then remains close to $0$~dB instead of converging to the performance attained with $\hat{J}=4$ or $5$. This empirical behavior is consistent with the NMSE and BLER results in Fig.~\ref{fig:J_performance}: mild overestimation is tolerable, whereas underestimation can prevent reliable target-channel recovery.

\subsection{Computational Complexity and Runtime}

For JCEIS-SVI, the dominant cost arises from RMP-based interference-channel inference. With $M$ reverse-diffusion steps, $D$ NGD updates, and $J$ inferred interference components, RMP uses approximately $MDJ$ score-network forward passes per transmission frame. STMP adds $MK$ score-network evaluations for the $K$ target channels, while each RMP update also includes an LMMSE interference-symbol update over the pilot resource elements. The interference score prior is implemented using the NCSN++ backbone~\cite{song2021scorebased}. It contains $39.47$ million parameters and requires $163.65$ giga floating-point operations (GFLOPs), for one forward pass with an input of size $2\times48\times256$.

Table~\ref{tab:runtime_complexity} reports measured channel-estimation runtime per realization for the proposed method and representative CE baselines on the indicated CPU or GPU platforms. It does not represent end-to-end receiver latency, since channel decoding and other data-detection operations are not included. The proposed method is substantially more computationally demanding than conventional linear CE methods and should therefore be regarded as a performance-oriented blind interference-suppression receiver rather than a low-complexity solution.

Table~\ref{tab:module_runtime} shows that RMP accounts for $80.15$~s of the $85.25$~s total runtime, including $78.21$~s for the NGD update and $1.94$~s for the embedded LMMSE symbol update. The runtime is thus dominated by score-prior evaluation. As indicated by the sensitivity results in Sec.~VI-H, practical cost reduction can be obtained by selecting smaller $M$ or $D$ values when a moderate performance loss is acceptable. Further reductions may be achieved through lighter score-network backbones, GPU batching, and accelerated generative inference~\cite{song2023consistency}.

\begin{table}[!t]
\caption{Channel-Estimation Runtime Per Realization}\label{tab:runtime_complexity}
\centering
\footnotesize
\begin{tabular}{lcc}
\toprule
\textbf{Algorithm} & \textbf{Runtime (s)} & \textbf{Platform}\\
\midrule
1D-LMMSE & 0.0050 & CPU\\
G-LMMSE & 4.4490 & CPU\\
ARA & 0.0380 & GPU\\
VBI-DSAE & 0.1620 & GPU\\
JCEIS-SVI & 85.2500 & GPU\\
VALSE & 351.4600 & CPU\\
\bottomrule
\end{tabular}
\end{table}

\begin{table}[!t]
\caption{Module-Level  Runtime of JCEIS-SVI}
\label{tab:module_runtime}
\centering
\footnotesize
\begin{tabular}{lc}
\toprule
\textbf{Module} & \textbf{Runtime (s)}\\
\midrule
STMP & 5.10\\
RMP--NGD & 78.21\\
Embedded LMMSE symbol update & 1.94\\
\midrule
\textbf{Total runtime} & \textbf{85.25}\\
\bottomrule
\end{tabular}
\end{table}

\section{Conclusion}\label{sec:conclusion}
This paper developed a joint channel estimation and interference suppression framework for MIMO-OFDM systems under strong interference. By explicitly factorizing the interference into interference channels and symbols, the proposed method exploits the bilinear interference structure and enables the incorporation of score-based priors within a variational inference framework. Building on this formulation, we developed the JCEIS-SVI algorithm by combining STMP for target-channel estimation with RMP for interference inference, and further designed a subcarrier-level covariance reconstruction method for MMSE-IRC detection. Simulation results show that, among practical schemes, the proposed method substantially improves both channel estimation and data detection in strongly interference-limited regimes.

Future work will pursue two directions. First, we will further improve estimation and detection performance by developing more expressive yet robust channel priors and more effective inference updates, particularly for severe propagation-model and distribution mismatches. Second, we will reduce online computational cost through lighter score-network architectures, accelerated generative inference, and implementation-level optimization.

\appendices
    \section{Derivation of \eqref{eq:chain_rule}}\label{appendix:likelihood_score}
The likelihood score $\nabla_{\hat{\mathbf{G}}_{j}^{(0)}} \log p(\mathbf{Y}_{\mathcal{G}} | \hat{\mathcal{G}}^{(0)})$ can be expressed as
    \begin{align}
        &\nabla_{\mathbf{G}_{j}^{(m)}} \log p(\mathbf{Y}_{\mathcal{G}} | \hat{\mathcal{G}}^{(0)}) \nonumber \\
        &\stackrel{(a)}{=} \frac{1}{p(\mathbf{Y}_{\mathcal{G}} | \hat{\mathcal{G}}^{(0)})} \nabla_{\mathbf{G}_{j}^{(m)}} p(\mathbf{Y}_{\mathcal{G}} | \hat{\mathcal{G}}^{(0)}) \nonumber \\
        &\stackrel{(b)}{=} \frac{1}{p(\mathbf{Y}_{\mathcal{G}} | \hat{\mathcal{G}}^{(0)})} \nabla_{\mathbf{G}_{j}^{(m)}} \int p(\mathbf{Y}_{\mathcal{G}}, \mathcal{S} | \hat{\mathcal{G}}^{(0)}) \, \mathrm{d}\mathcal{S} \nonumber \\
        &\stackrel{(c)}{=} \int \frac{1}{p(\mathbf{Y}_{\mathcal{G}} | \hat{\mathcal{G}}^{(0)})} \left[ \nabla_{\mathbf{G}_{j}^{(m)}} p(\mathbf{Y}_{\mathcal{G}}, \mathcal{S} | \hat{\mathcal{G}}^{(0)}) \right] \, \mathrm{d}\mathcal{S} \nonumber \\
        &\stackrel{(d)}{=} \int \frac{p(\mathbf{Y}_{\mathcal{G}}, \mathcal{S} | \hat{\mathcal{G}}^{(0)})}{p(\mathbf{Y}_{\mathcal{G}} | \hat{\mathcal{G}}^{(0)})} \nabla_{\mathbf{G}_{j}^{(m)}} \log p(\mathbf{Y}_{\mathcal{G}}, \mathcal{S} | \hat{\mathcal{G}}^{(0)}) \, \mathrm{d}\mathcal{S} \nonumber \\
        &\stackrel{(e)}{=} \mathbb{E}_{\mathcal{S} | \mathbf{Y}_{\mathcal{G}}, \hat{\mathcal{G}}^{(0)}} \left[ \nabla_{\mathbf{G}_{j}^{(m)}} \log p(\mathbf{Y}_{\mathcal{G}} | \mathcal{S}, \hat{\mathcal{G}}^{(0)}) \right], \label{eq:fisher_identity_derivation}
    \end{align}
    where (a) follows from the log-derivative rule $\nabla p = p \nabla \log p$; (b) is the marginalization over the latent symbols $\mathcal{S}$; (c) results from the Leibniz integral rule, as the support of $\mathcal{S}$ is independent of the channel parameters; (d) applies the log-derivative rule; and (e) is obtained by substituting the posterior distribution $p(\mathcal{S} | \mathbf{Y}_{\mathcal{G}}, \hat{\mathcal{G}}^{(0)})$ and noting that the prior $p(\mathcal{S})$ is independent of $\hat{\mathbf{G}}_{j}^{(0)}$.
    Let $\mathbf{U}(\mathcal{S}) = \mathbf{Y}_{\mathcal{G}} - \sum_{j} \mathbf{S}_j \hat{\mathbf{G}}_{j}^{(0)}$. By substituting the likelihood model defined in~\eqref{eq:G_posterior_form},~\eqref{eq:fisher_identity_derivation} can be further expanded as
    \begin{align}
        &\mathbb{E}_{\mathcal{S} | \mathbf{Y}_{\mathcal{G}}, \hat{\mathcal{G}}^{(0)}} \left[ \nabla_{\mathbf{G}_{j}^{(m)}} \log p(\mathbf{Y}_{\mathcal{G}} | \mathcal{S} , \hat{\mathcal{G}}^{(0)}) \right] \nonumber \\
        &= - \frac{1}{\sigma^2_{\mathcal{G}}} \nabla_{\mathbf{G}_{j}^{(m)}} \mathbb{E}_{\mathcal{S}  | \mathbf{Y}_{\mathcal{G}}, \hat{\mathcal{G}}^{(0)}} \left[ \left\| \mathbf{U}(\mathcal{S})\right\|_F^2 \right] \nonumber \\
        &= - \frac{1}{\sigma^2_{\mathcal{G}}} \nabla_{\mathbf{G}_{j}^{(m)}} \operatorname{Tr} \left( \mathbb{E}_{\mathcal{S}  | \mathbf{Y}_{\mathcal{G}}, \hat{\mathcal{G}}^{(0)}}  \left[ \left( \mathbf{U}(\mathcal{S}) \right)^{\mathsf{H}} \mathbf{U}(\mathcal{S}) \right] \right) \nonumber \\
        &=- \frac{1}{\sigma^2_{\mathcal{G}}} \nabla_{\mathbf{G}_{j}^{(m)}} \left( \left\| \mathbf{Y}_{\mathcal{G}} - \sum_{i} \hat{\mathbf{S}}_i \hat{\mathbf{G}}_{i}^{(0)} \right\|_F^2 \right. \nonumber \\
        &\quad \left. + \sum_{u,i} \operatorname{Tr}\left( {(\hat{\mathbf{G}}_{u}^{(0)})}^{\mathsf{H}} \mathbf{\Omega}_{ui} \hat{\mathbf{G}}_{i}^{(0)} \right) \right)
    \end{align}
    where $\hat{\mathbf{S}}_i = \mathbb{E}[\mathbf{S}_i | \mathbf{Y}_{\mathcal{G}}, \hat{\mathcal{G}}^{(0)}]$ denotes the posterior mean  and $\mathbf{\Omega}_{ui} =  \mathbb{E}[(\mathbf{S}_u - \hat{\mathbf{S}}_u)^{\mathsf{H}}(\mathbf{S}_i - \hat{\mathbf{S}}_i) | \mathbf{Y}_{\mathcal{G}}, \hat{\mathcal{G}}^{(0)}]= \mathbb{E}[\mathbf{S}_u^{\mathsf{H}} \mathbf{S}_i| \mathbf{Y}_{\mathcal{G}}, \hat{\mathcal{G}}^{(0)}] - \hat{\mathbf{S}}_u^{\mathsf{H}} \hat{\mathbf{S}}_i $ represents the posterior error covariance between the $u$-th and $i$-th symbol streams.

% \let\oldthebibliography\thebibliography
% \renewcommand{\thebibliography}[1]{\oldthebibliography{#1}\fontsize{6.5}{7.2}\selectfont}
% \IEEEtriggercmd{\enlargethispage{4\baselineskip}}
\bibliographystyle{IEEEtran}
\bibliography{reference}

\end{document}